\documentclass[useAMS,usenatbib,onecolumn
]{mnras}

\usepackage[usenames,dvipsnames]{xcolor}

\usepackage{amssymb, amsmath}
\usepackage{epsf}
\usepackage{bm}
\usepackage{ulem}

\usepackage{natbib}
\usepackage{graphicx}
\usepackage{cancel}

\def\la{\; \raise0.3ex\hbox{$<$\kern-0.75em\raise-1.1ex\hbox{$\sim$}}\;}
\def\ga{\;  \raise0.3ex\hbox{$>$\kern-0.75em\raise-1.1ex\hbox{$\sim$}}\;}

\title[Temperature-dependent r-modes in superfluid neutron stars stratified by muons]
{Temperature-dependent r-modes in superfluid neutron stars stratified by muons}

\author[E. M. Kantor, M. E. Gusakov]
{E.~M.~Kantor$^{1}$  \thanks{kantor@mail.ioffe.ru},
M. E. Gusakov$^{1}$ 
\\
$^1$Ioffe Physical-Technical Institute of the Russian Academy of
Sciences,
Polytekhnicheskaya 26, 194021 St.-Petersburg, Russia
}

\begin{document}

\date{Accepted 2017 xxxx. Received 2017 xxxx;
in original form 2017 xxxx}

\pagerange{\pageref{firstpage}--\pageref{lastpage}} \pubyear{2017}

\maketitle

\label{firstpage}


\begin{abstract}
We calculate the finite-temperature r-mode spectrum 
of a slowly rotating superfluid Newtonian neutron star 
neglecting the entrainment between neutron and proton liquid components
(i.e., neglecting the off-diagonal element of the entrainment matrix).
We show that 
for `minimal' NS core composition 
(neutrons, protons, and electrons)
only two $m=2$ r-modes exist --- normal mode, 
which is similar to ordinary r-mode in a nonsuperfluid star, 
and a superfluid temperature-dependent mode.
Accounting for muons in the core 
dramatically modifies
the oscillation spectrum,
resulting in an infinite 
set of superfluid r-modes, whose frequencies vary with temperature.
We demonstrate that the normal r-mode can exhibit avoided crossings 
with superfluid modes
at certain `resonance' temperatures, 
where it dissipates strongly,
which leads to substantial suppression of the r-mode instability near these temperatures.
The corresponding instability windows are calculated and discussed.
\end{abstract}


\maketitle

\section{Introduction}
\label{intro}

Since it had been shown that r-modes 
in neutron stars (NSs) 
can be unstable (\citealt*{andersson98,fm98}) 
and can effectively 
spin NSs down 
by transmitting 
their
angular momentum 
to gravitational waves, 
a significant progress has been made in understanding 
the r-mode physics 
(\citealt*{levin99,lm00,ac01,ak01,heyl02,ly03,hah11,hdh12,
ms13,as14,gck14a,gck14b,haskell15}). 
However, a number of important questions related 
to r-modes 
has remained unanswered.
The main problem
concerns the 
interpretation of observations
of hot rapidly rotating NSs 
in low-mass X-ray binary (LMXB) systems.
According to theoretical predictions 
the probability to observe such objects should be negligible
since they
should be heated and spun down 
by developing r-mode instability 
on a time-scale much smaller than the typical age 
of an LMXB system (\citealt{levin99}).
A number of possible solutions to this problem 
has been proposed 
(see, e.g., \citealt{haskell15} for a recent review). 
One of them (\citealt{gck14a,gck14b}) 
appeals to baryon superfluidity in the NS core.
The idea is
that the instability may be damped out
at some stellar temperatures 
by the resonance interaction 
of the normal r-mode with superfluid inertial modes 
\footnote{Following \cite{gck14a}, 
by the normal r-modes of a superfluid NS we mean the 
modes which, at a given temperature, 
have properties similar to 
ordinary r-modes in the non-superfluid NSs.
They correspond to co-motion of 
superfluid and normal liquid components,
and their eigenfrequencies do not depend on temperature. 
Normal r-modes are most unstable.
By superfluid inertial modes we 
understand inertial modes which show superfluid-like behaviour 
at a given temperature.
The 
oscillations 
of such modes
are mostly counter-moving, 
baryon current is almost zero, 
their eigenfrequencies depend on temperature. 
These modes have no analogue in non-superfluid NSs.
Note that,
with temperature variation, 
a given mode can switch from a `superfluid' regime 
to  `normal' regime
and vice versa,
experiencing avoided-crossings with neighbouring modes.}
in the vicinity of their avoided crossings. 
This mechanism was previously analysed qualitatively, 
but not quantitatively (\citealt{gck14a,gck14b,cgk14,kgc16,cgk17}). 
To study the resonance interaction of normal r-mode 
with superfluid modes one needs to calculate 
the temperature-dependent oscillation spectrum 
of a rotating superfluid neutron star.
This has not been done so far since 
most of
the studies 
have been concentrated  
on the limiting case of vanishing stellar temperature (e.g., \citealt{lm00,ac01,ak01,ly03,agh09,hap09}).

In this paper we present the first 
calculation
of the temperature-dependent spectrum of a rotating superfluid NS. 
The calculations are performed for two NS models:
The first one is standard (see, e.g., \citealt{ly03})
and assumes  minimal composition of the NS core 
(neutrons, protons, and electrons), 
while 
the second one
accounts for admixture of muons.
The presence of muons `stratifies' the NS core (\citealt*{kg14,dg16,pah16}),
which, as we show,  
has a dramatic effect on the low-frequency spectrum of rotating NSs.
Using these results, we study the effect of resonance interaction of normal r-mode
with superfluid modes on the stability of NSs
and plot
the `instability windows' (the regions of stellar rotation frequency and temperature, in which NSs are unstable, see, e.g., \citealt{ak01}) accounting for such interaction.

To simplify calculations we shall work in the 
Newtonian limit of the 
relativistic hydrodynamics, 
describing superfluid mixtures at a finite temperature
(see, e.g., \citealt*{ml91,ga06,gkcg13,gusakov16,gd16}). 
We also use a number of approximations. 
First, we adopt the Cowling approximation (i.e., gravitational potential is not perturbed, \citealt*{cowling41}).
Second, we consider only slowly rotating NSs,
expanding the corresponding eigenmodes in the Taylor series in the rotation frequency, 
and allowing only for the first two terms in the expansion.
Third, 
we assume that there is no entrainment, 
i.e., we neglect the off-diagonal element $Y_{\rm np}$ 
of the entrainment matrix (see \citealt*{ab76,gh05,ga06,gkh09a,gkh09b,ghk14}), $Y_{\rm np}=0$. 
This assumption is not unreasonable since numerical calculations 
show that entrainment is indeed small (\citealt{gkh09a,gkh09b,ghk14}).
Moreover, in the region where protons are non-superfluid, 
$Y_{\rm np}$ is strictly zero (see, e.g., \citealt{ghk14}).
We plan to relax this simplifying assumption 
in the subsequent publication.
And finally, we calculate only those modes 
which are purely toroidal to the leading order in rotation.

The paper is organized as follows.
In section \ref{equations} 
we briefly discuss the hydrodynamics of superfluid mixtures 
(section \ref{eqA})
and apply it to derive 
the system of equations
describing NS toroidal oscillations (section \ref{eqB}). 
Together with the
boundary conditions (section \ref{bc}),
this system
is solved numerically 
for a physics input 
from section \ref{PhysInput}.
The numerical results are presented in section \ref{res} 
for two models of the NS core with $\rm npe$ (section \ref{resI}) 
and $\rm npe\mu$ compositions (section \ref{resII}).
The oscillation spectra for these two models are compared and
interpreted
in section \ref{diff}.
In section \ref{damp} we apply the calculated spectra to study
damping of r-modes 
in the presence of avoided crossings 
of modes
and 
plot
the corresponding instability windows.
Finally, in section \ref{discus} 
we discuss our results 
and try to understand
how
the oscillation spectrum 
will be modified in a more realistic situation
of
non-zero entrainment. 

The paper also contains four appendices,
which
include some details on the derivation of the oscillation equations 
(Appendix \ref{derivation}), 
an analytical solution for the toroidal modes of a superfluid NS with $\rm npe$ core composition 
(Appendix \ref{npe}), 
the expression for the mutual friction coefficient 
(Appendix \ref{mf}),
and a list of typos in the illuminating paper by \cite*{pbr81}, whose analysis of r-modes in normal stars is similar to that presented here (Appendix \ref{typos}).

\section{Equations governing oscillations of a rotating NS}
\label{equations}

\subsection{General equations of superfluid hydrodynamics}
\label{eqA}

In our analysis we shall use 
the Newtonian limit of the relativistic hydrodynamics,  
formulated in \cite{gusakov16,gd16}
and describing
rotating superfluid mixture at a finite temperature. 
Below we formulate the main equations of this hydrodynamics.
We assume that the mixture is composed of nucleons (neutrons and protons),
which can be in superfluid state, and of normal leptons (electrons and muons). 
We ignore all the dissipative effects except for mutual friction, 
and shear viscosity, 
which turn out to be important for calculating the instability windows in section \ref{damp}.
We also neglect the vortex-related contribution to the energy density,
which is an accurate approximation for our problem (\citealt*{sauls89}), 
and do not consider the magnetic field effects.
The equations governing superfluid dynamics consist of:

(i) Energy-momentum conservation law
\begin{eqnarray}
T^{\mu\nu}_{;\nu}=0
\end{eqnarray}
with the energy-momentum tensor
\begin{eqnarray}
T^{\mu\nu}= (P+\epsilon) \, u^{\mu} u^{\nu} + P \, g^{\mu \nu} 
+ Y_{ik} \left( w^{\mu}_{(i)} w^{\nu}_{(k)} + \mu_i \, w^{\mu}_{(k)} u^{\nu} 
+ \mu_k \, w^{\nu}_{(i)} u^{\mu} \right)+\tau^{\mu\nu}, 
\label{Tmunufluid}
\end{eqnarray}
where $P$ is the pressure, 
$\epsilon$ is the energy density;
$g^{\mu\nu}$ is the metric tensor; 
small dissipative terms due to shear viscosity are contained in the tensor $\tau^{\mu\nu}$, see \cite{gkcg13} for details;
$Y_{ik}$ is the relativistic symmetric entrainment matrix (\citealt{ga06, gkh09a,gkh09b,ghk14}), 
analogue of the superfluid or mass-density matrix $\rho_{ik}$ 
of the non-relativistic theory (\citealt{ab76, bjk96, gh05, ch06, gusakov10}).
Here and below indices $i,k$ run over neutrons and protons: $i$, $k={\rm n}$, ${\rm p}$. 
Note that, 
in Eq.\ (\ref{Tmunufluid}) and further in the text we assume 
summation over the repeated indices $i$ and $k$.
Finally, $u^{\mu}$ is the four-velocity of the non-superfluid component 
(leptons and baryon thermal excitations), 
normalized by the condition $u_{\mu}u^{\mu}=-1$, 
and $w^{\nu}_{(k)}$ ($k={\rm n},{\rm p}$) is the four-vector 
that describes superfluid degrees of freedom 
(in particular, it is related to the superfluid velocity $v_{({\rm s}k)}^\nu$ by 
$v_{({\rm s}k)}^\nu=(w^{\nu}_{(k)}+\mu_k u^\nu)/m_k$, 
where $\mu_k$ and $m_k$ are, respectively, the relativistic chemical potential 
and the bare mass of particle species $k$).

(ii) Continuity equations for all particle species $j$
\begin{equation}
\partial_{\mu}j^{\mu}_{(j)}=0,
\end{equation}
where the particle current density for neutrons and protons is
\begin{equation}
j^{\mu}_{(k)}=n_k u^\mu+Y_{ik}w^\mu_{(i)},
\label{j1}
\end{equation}
and for leptons (electrons and muons)
\begin{equation}
j^{\mu}_{({\rm l})}=n_{{\rm l}} u^\mu
\label{j2}
\end{equation}
(here and below ${\rm l}={\rm e}$, ${\rm \mu}$).
In equations\ (\ref{j1}) and (\ref{j2})
$n_j$ is the number density for particle species~$j$.

(iii) Neutron `superfluid' equation 
\begin{eqnarray}
u_\nu 
\mathcal{V}^{\mu\nu}=\mu_{\rm n} n_{\rm n} f^\mu \label{sfl_gen}
\end{eqnarray}
where $\mathcal{V}^{\mu\nu}\equiv \partial^\mu(w_{({\rm n})}^\nu+\mu_{\rm n} u^\nu)-\partial^\nu(w_{({\rm n})}^\mu+\mu_{\rm n} u^\mu)$, 
and $f^\mu$ is given 
by the expression
\begin{eqnarray}
f^\mu=\alpha \perp^{\mu\nu} 
\mathcal{V}_{\nu\lambda}
\frac{1}{n_{\rm n}}Y_{{\rm n}k}w_{(k)\,\delta}\perp^{\lambda\delta} 
+ \frac{\beta_{\rm MF}}{\mathcal{V}_{({\rm M})}}  \perp^{\mu\eta} \perp^{\nu\sigma}  
\mathcal{V}_{\eta\sigma}
\mathcal{V}_{\lambda\nu}
\,\frac{1}{n_{\rm n}}Y_{{\rm n}k}w_{(k)\,\delta} \perp^{\lambda \delta}, \label{f_gen}
\end{eqnarray}
in which 
$\perp^{\mu\nu} \equiv g^{\mu\nu}+u^\mu u^\nu$, $\mathcal{V}_{({\rm M})} \equiv \sqrt{\mathcal{V}^{\mu}_{({\rm M})}\mathcal{V}_{({\rm M})\mu}}$, $\mathcal{V}^{\mu}_{({\rm M})} \equiv  
\frac{1}{2} \, \epsilon^{\mu \nu \gamma \delta} \, u_{\nu} \, \mathcal{V}_{\gamma \delta}$ (here $\epsilon^{\mu \nu \gamma \delta}$ is the Levi-Civita tensor, $\epsilon^{0123}=1$).
Kinetic coefficients $\alpha$ and $\beta_{\rm MF}$ depend on 
the details of interaction between the non-superfluid component and vortices. 
In what follows we will be interested in the so-called weak-drag regime, 
when vortices weakly interact with the normal component 
(this regime takes place in the NS cores, see, e.g., \citealt{mendell91,asc06}).
In this limit $\alpha=-1/(\mu_{\rm n} \mu_k Y_{{\rm n}k})$, 
while the coefficient $\beta_{\rm MF}$ is small and describes 
dissipation due to mutual friction 
(the dissipative mechanism related to scattering of electrons off the vortices).

These equations should be supplemented by the condition
ensuring that all the thermodynamic quantities are defined 
in the comoving frame in which $u^{\mu}=(1,\, 0,\, 0,\, 0)$,
\begin{equation}
u_{\mu}w^{\mu}_{(i)}=0,
\label{uw}
\end{equation}
as well as by the second law of thermodynamics,
\begin{equation}
d \epsilon = T \, d S + \sum_{j={\rm n,p,e,\mu}}\mu_j \, d n_j 
+ { Y_{ik} \over 2} \, d \left( w^{\alpha}_{(i)} w_{(k) \alpha} \right),
\label{2ndlaw}
\end{equation}
the expression for pressure,
\begin{equation}
P = -\epsilon +\sum_{j={\rm n,p,e,\mu}}\mu_j n_j + TS,
\label{pres}
\end{equation}
and the conditions 
\begin{gather}
n_{\rm p}=n_{\rm e}+n_{\mu},
\label{quasi}\\
Y_{{\rm p}k} w^{\mu}_{(k)} =0,
\label{cond1}
\end{gather}
which are always satisfied (\citealt{mendell91})
for the low-frequency hydrodynamic oscillations we are interested in;
these conditions indicate that protons are effectively locked to electrons and muons
by the electromagnetic forces.
In equations (\ref{2ndlaw}) and (\ref{pres}) $T$ is the temperature and 
$S$ is the entropy density.

\subsection{Oscillation equations}
\label{eqB}

Using the Newtonian limit of the above equations, 
let us consider small oscillations of a slowly rotating 
(with the spin frequency $\Omega$) 
non-dissipative Newtonian NS in the Cowling approximation. 
In what follows we shall allow for muons in the inner layers of NSs, 
assuming $\rm npe\mu$-composition, 
and also take into account possible
superfluidity of baryons (neutrons and protons) in the core.
Let all the quantities depend on time $t$ as ${\rm e}^{\imath\sigma t}$ 
in the coordinate frame rotating with the star. 
Then the linearised equations governing small oscillations of superfluid (hereafter SFL) NSs 
in that frame  
consist of:

(i) Euler equation
\begin{eqnarray}
-\sigma^2 {{\pmb \xi}_{\rm b}}+2 \imath \sigma {\pmb \Omega}\times {{\pmb \xi}_{\rm b}}=\frac{\delta w}{w_0^2}{\pmb \nabla} P_0-\frac{{\pmb \nabla} \delta P}{w_0}, 
\label{euler}
\end{eqnarray}
where $w=(P+\epsilon)/c^2$, 
$c$ is speed of light. 
Here and hereafter, the subscript $0$ denotes the equilibrium value
of some quantity (e.g., $P_0$) and 
$\delta$ stands for its Euler perturbation (e.g., $\delta P$). 
Finally, 
${\pmb \xi}_{\rm b}$ in equation (\ref{euler}) is the Lagrangian displacement of baryons, 
it is defined as
\begin{eqnarray}
{\pmb \xi}_{\rm b}\equiv \frac{{\pmb j}_{\rm b}}{\imath \sigma n_{\rm b}},
\end{eqnarray}
where $n_{\rm b}\equiv n_{\rm n}+n_{\rm p}$ and ${\pmb j}_{\rm b}\equiv {\pmb j}_{\rm n}+{\pmb j}_{\rm p}$ are the baryon number density and baryon current density, respectively.

(ii) Continuity equations for baryons and leptons (electrons and muons)
\begin{eqnarray}
\delta n_{\rm b}+{\rm div}(n_{\rm b} {\pmb \xi}_{\rm b})=0 \label{cont b}, \\
\delta n_{\rm l}+{\rm div}(n_{\rm l} {\pmb \xi})=0 \label{cont l}.
\end{eqnarray}
Here ${\pmb \xi}\equiv {\pmb j}_{\rm e}/(\imath \sigma n_{\rm e})$ is the Lagrangian displacement 
of the normal liquid component 
[we assume that all the normal-matter constituents 
(i.e., leptons and baryon thermal excitations)
move with one and the same normal velocity 
due to efficient particle collisions].
If neutrons are non-superfluid, then ${\pmb \xi}={\pmb \xi}_{\rm b}$
and hydrodynamic equations become essentially the same 
as in the normal matter (even if protons are SFL, 
see equations \ref{quasi} and \ref{cond1}), see e.g. \citealt{ga06}. 
Taking this into account, we, for brevity, 
shall call `normal' (or `non-superfluid') the liquid 
with non-superfluid neutrons,
irrespective of the actual state of protons.

(iii) The `superfluid' equation, analogue of the Euler equation 
for superfluid (neutron) liquid component
\begin{eqnarray}
h \sigma^2 {\pmb z}+ 2\imath \sigma {\pmb \Omega} \times {\pmb z}\,n_{\rm b} \mu_{\rm n}\left(1+\alpha \frac{n_{\rm b} \mu_{\rm n}}{c^2}\right)=
h \sigma^2 {\pmb z}- 2\imath h_1 \sigma {\pmb \Omega} \times {\pmb z}=c^2 n_{\rm e} {\pmb \nabla} \Delta \mu_{\rm e}+c^2 n_{\rm \mu} {\pmb \nabla} \Delta \mu_{\rm \mu},
\label{sfl1}
\end{eqnarray}
where ${\pmb z}\equiv{\pmb \xi}_{\rm b}-{\pmb \xi}$ is the superfluid Lagrangian displacement;
and
$\Delta \mu_{\rm l}\equiv \mu_{\rm n}-\mu_{\rm p}-\mu_{\rm l}$ is the chemical potential imbalance (note that
in equilibrium $\Delta \mu_{\rm l}=0$, see \citealt{hpy07}, thus $\delta \Delta \mu_{\rm l}=\Delta \mu_{\rm l}$).
Further,
\begin{eqnarray}
h=n_{\rm b} \mu_{\rm n} y, \label{beta}\\
h_1=\mu_{\rm n} n_{\rm b}\left(\frac{n_{\rm b}}{Y_{{\rm n}k}\mu_k}-1\right), 
\label{gamma} \\
y=\frac{n_{\rm b}Y_{{\rm pp}}}{\mu_{\rm n}(Y_{{\rm nn}}Y_{{\rm pp}}-Y_{{\rm np}}^2)}-1. \label{y}
\end{eqnarray}
Note that in equation (\ref{gamma}) summation over $k={\rm n}$, ${\rm p}$ is assumed.
Superfluid equation in the form (\ref{sfl1}) 
is valid in the weak-drag regime only, 
when the interaction between the neutron vortices and normal component (e.g., electrons) is weak, 
which is a typical situation in NSs 
(see, e.g., \citealt{mendell91,asc06}).
In addition, equation (\ref{sfl1}) assumes that Newtonian limit is justified.
Since in that limit the redshift ${\rm e}^{\nu/2}\approx 1$ 
and $\nu'/2=\nabla P_0/(w_0 c^2)\ll 1/R$ 
($R$ is the stellar radius, which is a typical length-scale of the problem; 
we also used the fact that $P_0\ll \epsilon_0$ in the Newtonian limit), 
we skipped all redshifts in equation (\ref{sfl1}), 
replacing, e.g.,
$\Delta \mu_{\rm l}^\infty\equiv \Delta\mu_{\rm l} {\rm e}^{\nu/2}$ 
with $\Delta \mu_{\rm l}$
in the right-hand side of this equation.
Moreover, the Newtonian limit implies that one can neglect
(i) the frame drag effect, which 
is $\sim \Omega_0^2 R^3 \Omega/c^3$, 
where $\Omega_0\equiv \left(G M/R^3 \right)^{1/2}$ is of the order of the Kepler frequency 
(and $M$ is the stellar mass);
(ii) all the terms $\sim \Omega^2 R^2/c^2$, since $c \rightarrow \infty$ in the Newtonian limit. 
Although for NSs it is not a well justified approximation 
(because 
they are essentially relativistic objects),
however, it generally gives qualitatively correct results (see, e.g. \citealt*{ioj15}).%
%
\footnote{A detailed discussion of r-modes in relativistic neutron stars
and the related problems (e.g., the problem of a continuous r-mode spectrum
in non-barotropic relativistic stars)
can be found in \cite*{kojima98,kh99,laf01, lfa03,yl02,yl03, law04}. }
%
%
The above equations should be supplemented with the `equation of state' (EOS),
\begin{eqnarray}
\delta n_i=\frac{\partial n_i}{\partial P} \delta P+\frac{\partial n_i}{\partial \Delta \mu_{\rm e}} \Delta \mu_{\rm e}+\frac{\partial n_i}{\partial \Delta \mu_{\rm \mu}} \Delta \mu_{\rm \mu}. \label{eos}
\end{eqnarray}
In what follows we shall use 
$P$, $\Delta \mu_{\rm e}$ and $\Delta \mu_{\rm \mu}$ as independent thermodynamic variables.
We shall consider a slowly rotating NS with $\Omega\ll \Omega_0$, 
and shall expand all the relevant quantities in a power series 
in small parameter $\Omega/\Omega_0$, 
keeping the first two terms in the expansion.
This means, that the NS rotational oblateness, 
that has an order $(\Omega/\Omega_0)^2$, 
should be accounted for.
The isobaric surfaces of a rotating NS 
are not spherical as in non-rotating NSs. 
When the rotation is slow, the coordinates $r$ and $\theta$ 
of the isobaric surfaces
are related by (\citealt*{chandra33,cr63,hartle67,pbr81,saio82})
\begin{eqnarray}
r=x\left[1-\Omega^2 \alpha(x) {\rm cos}^2 \theta\right], 
\label{shape}
\end{eqnarray}
where $r$ and $\theta$ are the radial coordinate 
and polar angle in the spherical coordinate system 
with the origin at the stellar centre
and the axis z parallel to ${\pmb \Omega}$.
Every isobaric surface is characterized by the $\theta$-independent parameter $x$, $x={\rm const}$. 
The function $\alpha(x)$ (not to be confused with the mutual friction coefficient $\alpha$, see equation \ref{f_gen}) is to be determined from the hydrostatic equilibrium equations. 
To calculate $\alpha(x)$ we used the Hartle scheme (\citealt*{hartle67,ht68}). 
It allows us to find the structure of the relativistic 
slowly rotating star up to (and including) the terms $\sim (\Omega/\Omega_0)^2$.
In what follows it turns out to be more convenient to work in the coordinates
$x$, $\theta$ and $\phi$ instead of $r$, $\theta$ and $\phi$ ($\phi$ is the azimuthal angle).
Then the oblateness appears in the oscillation equations through the function $\alpha(x)$.

In the present paper we are interested in 
the small-amplitude (linear) 
oscillations, 
which depend on time $t$ and azimuthal angle $\phi$ as 
${\rm e}^{\imath \sigma t+\imath m \phi}$ 
in the frame rotating with the star, 
and which have the eigenfrequencies $\sigma$ 
vanishing at $\Omega \rightarrow 0$. 
Thus, up to the terms $\sim (\Omega/\Omega_0)^2$,
$\sigma$ and the Euler perturbation 
of any (scalar) thermodynamic parameter $f$ 
(e.g., $P$, $\mu_{\rm l}$, $n_{\rm b}$, etc.)
can be presented as (e.g., \citealt*{pbr81,lf99,lm00})
\begin{eqnarray}
\sigma=\Omega \sigma_0\left(1+\Omega^2 \sigma_1\right),\\
\delta f=\Omega^2 \delta f_1 {\rm exp}(\imath \sigma t+\imath m \phi). 
\label{deltaf}
\end{eqnarray}

Concerning Lagrangian displacements ${\pmb \xi}_{\rm b}$, $\pmb \xi$, and $\pmb z$, 
we look for a purely toroidal (to leading order in $\Omega/\Omega_0$) 
oscillation modes which assume the following ordering (\citealt{lf99})
\begin{eqnarray}
d_r=\Omega^2 d_r^1 {\rm exp}(\imath \sigma t+\imath m \phi), 
\label{dr}\\
d_\theta=(d_\theta^0+\Omega^2 d_\theta^1) {\rm exp}(\imath \sigma t+\imath m \phi), 
\label{dth}\\
d_{\phi}=(d_\phi^0+\Omega^2 d_\phi^1) {\rm exp}(\imath \sigma t+\imath m \phi), 
\label{dph}
\end{eqnarray}
where ${\pmb d}$ stands for the displacements ${\pmb \xi}_{\rm b}$, ${\pmb \xi}$, or ${\pmb z}$. 
Then the continuity equations (\ref{cont b}) and (\ref{cont l}) for 
baryons ($\rm b$)
and for leptons ($\rm l=e$ or $\rm \mu$) 
can be presented, to the leading order in $\Omega/\Omega_0$, as
\begin{eqnarray}
\frac{\partial}{\partial \theta}{\rm sin}\theta \xi_{{\rm b}\theta}^0+\imath m \xi_{{\rm b}\phi}^0=0, \label{contb0} \\
\frac{\partial}{\partial \theta}{\rm sin}\theta \xi_\theta^0+\imath m \xi_\phi^0=0. 
\label{contl00}
\end{eqnarray}
Below we express all the perturbations through the displacements 
${\pmb \xi}_{\rm b}$ and ${\pmb z}$, 
and the functions $\delta P$, $\Delta \mu_{\rm e}$, and $\Delta \mu_{\rm \mu}$. 
Thus, it is convenient to rewrite 
equation (\ref{contl00}) with the help of equation (\ref{contb0}) 
and the definition ${\pmb z}={\pmb \xi}_{\rm b}-{\pmb \xi}$ as
\begin{eqnarray}
\frac{\partial}{\partial \theta}{\rm sin}\theta z_\theta^0+\imath m z_\phi^0=0. \label{contl0}
\end{eqnarray}
The Euler equation (\ref{euler}) and the superfluid equation (\ref{sfl1}) give us, to the leading order in $\Omega/\Omega_0$,
\begin{eqnarray}
\sigma_0 \xi_{{\rm b}\theta}^0+2\imath{\rm cos}\theta \xi_{{\rm b}\phi}^0=-\frac{\imath}{m}\frac{\partial}{\partial \theta}{\rm sin}\theta \left(\sigma_0 \xi_{{\rm b}\phi}^0-2\imath {\rm cos}\theta \xi_{{\rm b}\theta}^0\right), \label{Euler0} \\
h(x)\sigma_0 z_\theta^0+2\imath h_1(x){\rm cos}\theta z_\phi^0=-\frac{\imath}{m}\frac{\partial}{\partial \theta}{\rm sin}\theta \left[h(x)\sigma_0 z_\phi^0-2\imath h_1(x){\rm cos}\theta z_\theta^0\right].  \label{sfl0general}
\end{eqnarray}
These equations decouple into two systems, 
equations (\ref{contb0}), (\ref{Euler0}) 
and equations (\ref{contl0}), (\ref{sfl0general}) (\citealt{ac01,ly03,agh09}). 
Equations (\ref{contb0}) and (\ref{Euler0}) describe the normal r-modes, 
analogous to ordinary r-modes of non-superfluid NSs, 
while equations (\ref{contl0}) and (\ref{sfl0general}) 
describe superfluid modes driven by the relative motion 
(represented by the vector ${\pmb z}$)
of superfluid and normal (non-superfluid) liquid components. 
The solution to these two systems gives
the following formulas
for eigenfrequencies 
\begin{eqnarray}
\sigma_0=\frac{2m}{l(l+1)},
\label{sigmanorm}\\
\sigma_0=\frac{2m}{l(l+1)}\frac{h_1(x)}{h(x)}
\label{sigma0general}
\end{eqnarray}
and eigenfunctions
\begin{eqnarray}
\xi_{{\rm b}\theta}^0=\imath m C_{lm}(x)\frac{P_l^m({\rm cos}\theta)}{{\rm sin}\theta},\;\;\;\;\xi_{{\rm b}\phi}^0=- C_{lm}(x)\frac{d}{d\theta}P_l^m({\rm cos}\theta), \label{xib0}\\
z_\theta^0=\imath m C_{zlm}(x)\frac{P_l^m({\rm cos}\theta)}{{\rm sin}\theta},\;\;\;\;z_\phi^0=- C_{zlm}(x)\frac{d}{d\theta}P_l^m({\rm cos}\theta) \label{z0}
\end{eqnarray}
of normal and superfluid modes, respectively.
In equations (\ref{xib0}) and (\ref{z0}) 
$P_l^m({\rm cos}\theta)$ are the Legendre polynomials. 

Since the function $h_1(x)/h(x)$ in equation (\ref{sigma0general})
generally varies throughout the star, 
the frequency (\ref{sigma0general}) 
cannot be a global frequency for the star as a whole 
--- each layer has its own different eigenfrequency. 
This indicates that there are no purely toroidal superfluid modes in the system, 
an admixture of poloidal component is required.
In other words, for superfluid modes $z_r$ should be non-zero 
at $\Omega \rightarrow 0$, making them (mixed) inertial modes rather than r-modes. 
The same conclusion 
	has been arrived at
	in \cite{agh09}, 
although these authors obtained this result 
in the limit of vanishing stellar temperature ($T=0$).

However, if we assume that $Y_{np}=0$ then $h_1(x)=h(x)$ 
(see equations \ref{beta}--\ref{y}) 
and the superfluid equation (\ref{sfl1}) simplifies, 
\begin{eqnarray}
h(x)\left(\sigma^2 {\pmb z}-2\imath \sigma {\pmb \Omega} \times {\pmb z}\right)=c^2 n_{\rm e} {\pmb \nabla} \Delta \mu_{\rm e}+c^2 n_{\rm \mu} {\pmb \nabla} \Delta \mu_{\rm \mu}.
\label{sfl}
\end{eqnarray}
Then the eigenfrequency (\ref{sigma0general}) of superfluid modes, 
\begin{eqnarray}
\sigma_0=\frac{2m}{l(l+1)},
\end{eqnarray}
is independent of $x$ (\citealt{ac01,ly03,agh09}), becoming a global solution.
Consequently, purely toroidal (to leading order in $\Omega/\Omega_0$) 
superfluid modes are possible 
in the limit $Y_{\rm np}=0$.
Generally
$|Y_{\rm np}|\ll Y_{\rm pp},Y_{\rm nn}$ (see section \ref{PhysInput} for a numerical example);
also, when protons are non-superconducting, one has $Y_{\rm np}=0$. 
Thus, in this paper we restrict ourselves 
 to analysis of only
the
limiting  case of vanishing entrainment, $Y_{\rm np}=0$, 
which may be not too bad approximation,
see a corresponding discussion in section \ref{discus}.

We see that the eigenfrequencies of
all oscillation modes
with $\sigma\propto \Omega$ 
and vanishing radial displacement 
coincide 
(to the lowest order in $\Omega/\Omega_0$) at $Y_{\rm np}=0$, 
while their eigenfunctions depend 
on the functions $C_{lm}(x)$ and $C_{zlm}(x)$ 
(see equations \ref{xib0} and \ref{z0}).
To find these functions
we need to proceed to the next order in $\Omega/\Omega_0$. 
Then the following relation 
is useful 
(it follows from equation \ref{shape}),
\begin{equation}
\frac{\partial f(r,\theta)}{\partial \theta}\approx  \frac{\partial f(x,\theta)}{\partial \theta}-\frac{\partial f(x,\theta)}{\partial x} 2\Omega^2 x \alpha(x) {\rm cos} \theta {\rm sin}\theta. \label{changevar}
\end{equation}
Using it
along with expansions (\ref{deltaf})--(\ref{dph}), 
we find 
from the continuity equations for, respectively, baryons and leptons:
\\
(I) Baryons
\begin{eqnarray}
\frac{1}{n_{\rm b}} \left(\frac{\partial n_{\rm b}}{\partial P} \delta P_1+\frac{\partial n_{\rm b}}{\partial \Delta \mu_{\rm e}} \Delta \mu_{{\rm e}1}+\frac{\partial n_{\rm b}}{\partial \Delta \mu_{\rm \mu}} \Delta \mu_{{\rm \mu} 1} \right)=\nonumber \\ 
-\frac{1}{x^2 n_{\rm b}}\frac{\partial}{\partial x}(x^2 n_{\rm b} \xi_{{\rm b}r}^1)-\frac{1}{x {\rm sin} \theta}\left(\frac{\partial}{\partial \theta}{\rm sin}\theta \xi_{\rm b \theta}^1+\imath m\xi_{\rm b\phi}^1\right)-2\alpha(x){\rm cos}\theta{\rm sin}\theta\left(g\frac{w_0}{n_{\rm b}}\frac{\partial n_{\rm b}}{\partial P}-\frac{\partial}{\partial x}\right)\xi_{\rm b \theta}^0, \label{contb2}
\end{eqnarray}
(II) Leptons (${\rm l=e}$, ${\rm \mu}$) 
\begin{eqnarray}
\frac{1}{n_{\rm l}}\left(\frac{\partial n_{\rm l}}{\partial P} \delta P_1+\frac{\partial n_{\rm l}}{\partial \Delta \mu_{\rm e}} \Delta \mu_{{\rm e}1}+\frac{\partial n_{\rm l}}{\partial \Delta \mu_{\rm \mu}} \Delta \mu_{{\rm \mu} 1}\right)=\nonumber \\ 
-\frac{1}{n_{\rm l}x^2}\frac{\partial}{\partial x}(x^2 n_{\rm l} \xi_{{\rm b}r}^1)-\frac{1}{x {\rm sin} \theta}\left(\frac{\partial}{\partial \theta}{\rm sin}\theta \xi_{{\rm b}\theta}^1+\imath m\xi_{{\rm b}\phi}^1\right)-2\alpha(x){\rm cos}\theta{\rm sin}\theta \left(g\frac{w_0}{n_{\rm l}}\frac{\partial n_{\rm l}}{\partial P}-\frac{\partial}{\partial x}\right)\xi_{{\rm b}\theta}^0 \nonumber \\
+\frac{1}{n_{\rm l}x^2}\frac{\partial}{\partial x}\left(x^2 n_{\rm l} z_r^1\right)+\frac{1}{x {\rm sin} \theta}\left(\frac{\partial}{\partial \theta}{\rm sin}\theta z_{\theta}^1+\imath m z_{\phi}^1\right)+2\alpha(x){\rm cos}\theta{\rm sin}\theta \left(g\frac{w_0}{n_{\rm l}}\frac{\partial n_{\rm l}}{\partial P}-\frac{\partial}{\partial x}\right)z_{\theta}^0, \label{contl2}
\end{eqnarray}
where to calculate the thermodynamic derivatives in these equations
it is assumed that the independent variables 
are $P$, $\Delta \mu_{\rm e}$ and $\Delta \mu_{\rm \mu}$;
$g=-\nabla P_0/w_0$; and we used the fact that 
$\nabla n_{i0}=(\partial n_i/\partial P) \,\nabla P_0$ 
(we remind that the chemical potential imbalances vanish in equilibrium,
$\Delta \mu_{\rm l}=0$, see, e.g., \citealt{hpy07}).

The Euler and superfluid equations written up to the second order give 
(see Appendix \ref{derivation} for some details on the derivation of these equations):
\\
(III) $r$-component of the Euler equation
\begin{eqnarray}
2\imath \sigma_0 {\rm sin}\theta \xi_{\rm b \phi}^0=\frac{\partial }{\partial x}\frac{\delta P_1}{w_0}-\frac{\mu_{\rm n}}{w_0^2 c^2}\left(\frac{\partial n_{\rm b}}{\partial \Delta \mu_{\rm e}}\Delta \mu_{{\rm e} 1}+\frac{\partial n_{\rm b}}{\partial \Delta \mu_{\rm \mu}}\Delta \mu_{{\rm \mu} 1} \right) \frac{\partial P_0}{\partial x}, 
\end{eqnarray}
(IV) $\phi$-component of the Euler equation
\begin{eqnarray}
\sigma_0^2 \xi_{{\rm b}\phi}^0-2\imath \sigma_0 {\rm cos}\theta \xi_{{\rm b}\theta}^0=\frac{\imath m}{x\, {\rm sin}\theta\, w_0}\delta P_1, \label{deltaP}
\end{eqnarray}
(V) $r$-component of the superfluid equation
\begin{eqnarray}
2\imath h  \sigma_0 {\rm sin}\theta z_{\phi}^0=c^2 n_{\rm e} \frac{\partial \Delta \mu_{{\rm e} 1}}{\partial x}+c^2 n_{\rm \mu} \frac{\partial \Delta \mu_{{\rm \mu} 1}}{\partial x}, \label{sflr} 
\end{eqnarray}
(VI) $\phi$-component of the superfluid equation
\begin{eqnarray}
\sigma_0^2 z_{\phi}^0-2\imath \sigma_0 {\rm cos}\theta z_{\theta}^0=c^2\frac{\imath m}{x\, {\rm sin}\theta \, h}\left(n_{\rm e} \Delta \mu_{{\rm e} 1}+n_{\rm \mu} \Delta \mu_{{\rm \mu} 1}\right),  \label{deltamu} 
\end{eqnarray}
(VII) $\theta$-component of the Euler equation
\begin{eqnarray}
\sigma_0 \xi_{{\rm b}\theta}^1+2\imath {\rm cos}\theta \xi_{{\rm b}\phi}^1+2\sigma_1\left(\sigma_0 \xi_{{\rm b}\theta}^0+\imath{\rm cos}\theta \xi_{{\rm b}\phi}^0\right)\nonumber \\ 
=-\frac{\imath}{m}\frac{\partial}{\partial \theta}{\rm sin}\theta\left[\sigma_0 \xi_{{\rm b}\phi}^1-2\imath({\rm cos}\theta \xi_{{\rm b}\theta}^1+{\rm sin}\theta \xi_{{\rm b}r}^1)+2\sigma_1\left(\sigma_0 \xi_{{\rm b}\phi}^0-\imath{\rm cos}\theta \xi_{{\rm b}\theta}^0\right)\right]\nonumber \\
-\frac{2\imath}{m}{\rm sin}^2 \theta {\rm cos}\theta \alpha(x)\left[(\sigma_0+2m) \xi_{{\rm b}\phi}^0-2\imath {\rm cos}\theta \xi_{{\rm b}\theta}^0\right], \label{Eulertheta}
\end{eqnarray}
(VIII) $\theta$-component of the superfluid equation
\begin{eqnarray}
\sigma_0 z_\theta^1+2\imath {\rm cos}\theta z_\phi^1+2\sigma_1\left(\sigma_0 z_\theta^0+\imath{\rm cos}\theta z_\phi^0\right)\nonumber \\ =-\frac{\imath}{m}\frac{\partial}{\partial \theta}{\rm sin}\theta\left[\sigma_0 z_\phi^1-2\imath({\rm cos}\theta z_\theta^1+{\rm sin}\theta z_r^1)+2\sigma_1\left(\sigma_0 z_\phi^0-\imath{\rm cos}\theta z_\theta^0\right)\right]\nonumber \\
-\frac{2\imath}{m}{\rm sin}^2 \theta {\rm cos}\theta \alpha(x)\left[(\sigma_0+2m) z_{\phi}^0-2\imath {\rm cos}\theta z_{\theta}^0\right].
\label{sfltheta}
\end{eqnarray}

To solve this system, let us introduce 
the functions $\xi_{{\rm b}\theta}^1$, 
$\xi_{{\rm b}\phi}^1$, $z_\theta^1$, 
and $z_\phi^1$ as a sum of toroidal and poloidal components 
(e.g., \citealt{saio82})
\begin{eqnarray}
\xi_{{\rm b}\theta}^1=\frac{\partial}{\partial \theta}Q(x,\theta)+\frac{\imath m T(x,\theta)}{{\rm sin}\theta}, \\
\xi_{{\rm b}\phi}^1=\frac{\imath m Q(x,\theta)}{{\rm sin}\theta} -\frac{\partial}{\partial \theta}T(x,\theta), \\
z_\theta^1=\frac{\partial}{\partial \theta}Q_z(x,\theta)+\frac{\imath m T_z(x,\theta)}{{\rm sin}\theta}, \\
z_\phi^1=\frac{\imath m Q_z(x,\theta)}{{\rm sin}\theta} -\frac{\partial}{\partial \theta}T_z(x,\theta).
\end{eqnarray}
As in the normal (nonsuperfluid) stars (e.g., \citealt{lf99}),
we can now expand the functions 
$z_{r}^1(x,\theta)$, $\xi_{{\rm b} r}^1(x,\theta)$, 
$Q(x,\theta)$, $T(x,\theta)$, $Q_z(x,\theta)$, and $T_z(x,\theta)$, 
as well as $\delta P_1(x,\theta)$, $\Delta \mu_{\rm e 1}(x,\theta)$ and $\Delta \mu_{\rm \mu 1}(x,\theta)$, 
into Legendre polynomial series with fixed $m$,
\begin{eqnarray}
\xi_{{\rm b} r}^1(x,\theta)=\sum_{l_2}\xi_{{\rm b} r\, l_2m}^1(x) P_{l_2}^m(\rm cos\theta), \label{expxib}\\
z_{r}^1(x,\theta)=\sum_{l_2}z_{r\, l_2m}^1(x) P_{l_2}^m(\rm cos\theta), \\
Q(x,\theta)=\sum_{l_2}Q^1_{l_2m}(x) P_{l_2}^m(\rm cos\theta), \\
Q_z(x,\theta)=\sum_{l_2}Q^1_{z\,l_2m}(x) P_{l_2}^m(\rm cos\theta), \label{expQz} \\ 
T(x,\theta)=\sum_{l_1}T^1_{l_1m}(x) P_{l_1}^m(\rm cos\theta), \\
T_z(x,\theta)=\sum_{l_1}T^1_{z\,l_1m}(x) P_{l_1}^m(\rm cos\theta), \label{expRz} \\
\delta P_1(x,\theta)=\sum_{l_2}\delta P_{1\,l_2m}(x) P_{l_2}^m(\rm cos\theta), \label{expP1} \\
\delta \mu_{\rm e 1}(x,\theta)=\sum_{l_2}\delta \mu_{\rm e 1\,l_2m}(x) P_{l_2}^m(\rm cos\theta), \label{expmue1} \\
\delta \mu_{\rm \mu 1}(x,\theta)=\sum_{l_2}\delta \mu_{\rm \mu 1\,l_2m}(x) P_{l_2}^m(\rm cos\theta), \label{expmumu1}
\end{eqnarray}
where the summation goes over $l_2$ and $l_1$ 
($l_2=m+2k+1$ and $l_1=m+2k$, $k=0$, $1$, $2$, $\ldots$) and no summation over $m$ is assumed.

Substituting now these expansions, as well as expressions 
(\ref{xib0}) and (\ref{z0}) into equations (I)--(VIII) 
and equating coefficients 
at the same Legendre polynomials,
we first of all see that only the solutions with $l=m$ result in 
a non-contradictory system of equations 
(to see this, it is sufficient to consider equations (III)--(VI) only). 
Then it can be noted that only the terms 
in expansions (\ref{expxib})--(\ref{expQz}) and (\ref{expP1})--(\ref{expmumu1})
with $k=0$ 
contribute
to the terms proportional to the `lowest' Legendre polynomial in each equation, 
while the functions $T^1(x,\theta)$ and $T^1_z(x,\theta)$ completely drop out from those terms. 
The system of equations resulting from the lowest Legendre polynomials is thus closed,
and since in what follows we will be interested only in that system,
we can skip the summation in (\ref{expxib})--(\ref{expQz}), 
accounting only for the first terms there.
Other terms in the expansion 
[as well as the functions $T(x,\theta)$ and $T_z(x,\theta)$] 
cannot be constrained from equations (I)--(VIII);
to constrain them one needs to work in the next order in $\Omega/\Omega_0$.

Setting $l=m$, 
we obtain a system of four first-order differential equations 
for the functions $\xi_{{\rm b} r}^1(x)$, $z_{r}^1(x)$, $C(x)$, 
and $C_{z}(x)$ 
(from here on 
we skip the indices $lm$ 
and $l_2 m$)
of the form
\begin{eqnarray}
d\,'(x)=A(\sigma_1,x)d(x), \label{system}
\end{eqnarray}
where the vector 
$d(x) \equiv \left[\xi_{{\rm b} r}^1(x), z_{r}^1(x), 
C(x), C_{z}(x)\right]$; 
$A(\sigma_1,x)$ is a certain matrix depending on various thermodynamic functions, 
entrainment matrix coefficients, NS oblateness, pressure profile etc.
With the appropriate boundary conditions 
(see section \ref{bc}), the system (\ref{system}) 
constitute
the eigenvalue problem for $\sigma_1$. 
It is remarkable, that for  $\rm npe$-matter 
(when muons are absent) the solution to the system (\ref{system}), 
as well as the value of $\sigma_1$ 
can be 
found 
{\it analytically} (see Appendix \ref{npe}).

\section{Boundary conditions}
\label{bc}

At the stellar surface 
[$x=R+O(\Omega^2/\Omega_0^2$)]
we require that the Lagrangian perturbation of the pressure to be zero,
\begin{eqnarray}
\delta P(R)+P_0'(R)\xi_{{\rm b} r}^1(R)=0.
\end{eqnarray}
At the stellar centre ($x\rightarrow 0$) 
the system (\ref{system}) gives the following asymptotes 
for $\xi_{{\rm b} r}^1(x)$ and $C(x)$,
\begin{eqnarray}
\xi_{{\rm b} r}^1(x),\,C(x) \propto x^m,\\
\xi_{{\rm b} r}^1(x)=\frac{(1+m)\sigma_1}{1+2m}C(x).
\end{eqnarray}
Additionally, if neutrons in the centre are superfluid, one also has
\begin{eqnarray}
z_{r}^1(x),\,C_{z}(x) \propto x^m,\\
z_{r}^1(x)=\frac{(1+m)\sigma_1}{1+2m}C_z(x).
\end{eqnarray}

Analysis of equations (III)--(VI) requires
 the functions $C(x)$ and $C_z(x)$ to
be continuous throughout, respectively, 
the star and the superfluid region 
(in particular, there should be no discontinuities 
of $C(x)$ and $C_z(x)$ at the crust-core interface and at the threshold density 
where muons appear)
\footnote{We do not account for the density jumps in the crust and at the core-crust interface.}.
Next, the continuity equation for baryons, equation (I), 
requires the function $\xi_{{\rm b}r}^1(x)$ also to be continuous 
throughout
the star.
The continuity equation for leptons, equation (II), 
implies the continuity of $\xi_r^1(x)$. 
Thus, $z_r^1(x)$ should vanish at the superfluid/non-superfluid interface 
[since $z_r^1(x)=\xi_{{\rm b}r}^1(x)-\xi_{r}^1(x)$ 
and $z_r^1(x)=0$ in the non-superfluid region].
These boundary conditions together with the system (\ref{system}) constitute 
an eigenvalue problem for $\sigma_1$, 
which should be solved numerically.

\section{Physics input}
\label{PhysInput}

In our numerical calculations 
we adopt the \cite*{hh99} parametrization 
of APR equation of state (\citealt*{apr98}) 
for the matter in the core 
(at densities $\rho>\rho_{\rm cc}$) and  
the equation of state BSk20 (\citealt*{pfcpg13}) 
to describe the crust ($\rho<\rho_{\rm cc}$).
We choose $\rho_{\rm cc}\approx 1.30\times 10^{14}\,\rm g\, cm^{-3}$ 
to avoid discontinuity in the density profile (at this density both EOSs give the same pressure).
Notice, that the parametrization of \cite{hh99} 
allows for muons ($\rm npe\mu$-composition), 
that appear first at $n_{\rm b\mu}\approx 0.133\,\rm fm^{-3}$ 
($\rho_{\rm \mu}\approx 2.26\times 10^{14}\,\rm g\, cm^{-3}$).
To illustrate the effect of muons, 
in this paper
we consider 
two NS models: 
model I is simplified, muons are artificially switched off at any density;
model II is more realistic, 
it allows for the presence of muons at $\rho>\rho_{\rm \mu}$. 
The figures presented in this section are plotted for the model II.

All numerical 
examples considered 
in what follows (except for Fig.\ \ref{Fig:spectrumnpe})
are obtained for an NS with the mass $M=1.4M_{\odot}$. 
For the model II (accounting for muons) 
the circumferential radius for such a star is $R=12.1$~km,
the central density is 
$\rho_{\rm c}=9.48 \times 10^{14}$~g~cm$^{-3}$;
for the model I (muons are switched off) 
$R=12.2$~km, $\rho_{\rm c}=9.26 \times 10^{14}$~g~cm$^{-3}$. 

\begin{figure}
    \begin{center}
        \leavevmode
        \epsfxsize=3in \epsfbox{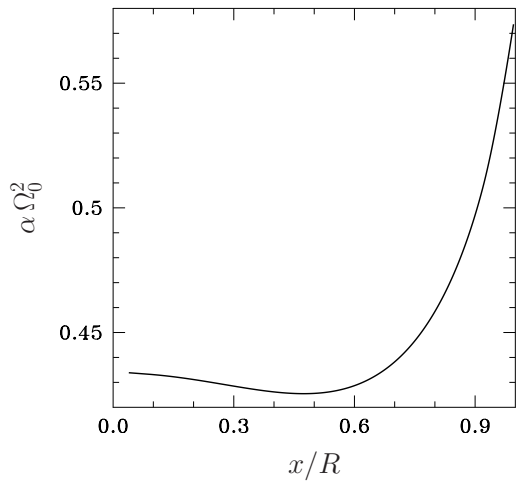}
    \end{center}
    \caption{$\alpha$ (in units of $\Omega_0^{-2}$) versus normalized radial coordinate $x/R$.
		}
    \label{Fig:alpha}
\end{figure}

To find the oblateness of a rotating NS 
we used the Hartle scheme (\citealt{hartle67,ht68}). 
The function $\alpha(x)$ (see equation \ref{shape}) 
that parametrizes the oblateness in oscillation equations 
is shown in Fig.\ \ref{Fig:alpha}. 
%
%
%
Note that, to solve the Newtonian oscillation equations we, 
somewhat inconsistently, 
used 
hydrostatic NS models, 
calculated in full General Relativity 
(for both rotating and non-rotating NS configurations); 
the EOS employed by us is also relativistic in a sense 
that $P \sim \epsilon$ 
(pressure is comparable to the energy density). 
One should bear in mind this inconsistency 
which can affect the results.
To make the calculations more reliable
it would be desirable 
to describe also NS oscillations 
within the fully relativistic framework.

\begin{figure}
    \begin{center}
        \leavevmode
        \epsfxsize=3in \epsfbox{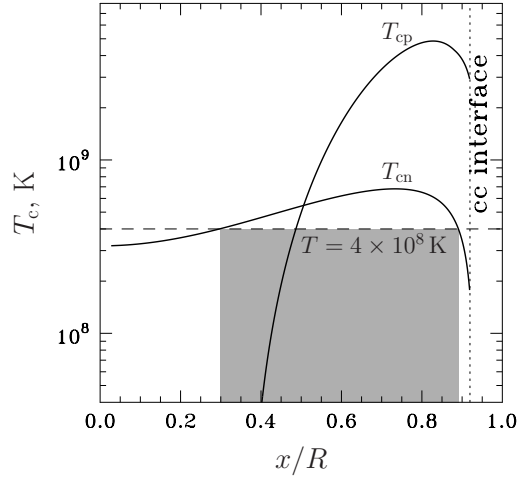}
    \end{center}
    \caption{Critical temperature profiles for neutrons and protons 
    	versus normalized radial coordinate $x$. 
    	The grey-filled region corresponds to the region of neutron superfluidity 
    	at $T=4\times 10^8\,\rm K$.
		}
    \label{Fig:Tc}
\end{figure}

When modelling the effects of superfluidity 
we allow for the neutron and proton superfluidity in the NS core, 
while neutron 
pairing
in the crust is neglected
(the crustal superfluidity should not affect strongly the global oscillations of NSs).
The adopted model of nucleon superfluidity is presented in Fig. \ref{Fig:Tc}, 
where the critical temperature profiles $T_{{\rm c}i}(x)$ for neutrons and protons
are shown 
as functions of $x$.
This model qualitatively does not contradict the results of
microscopic calculations  
(see, e.g., \citealt*{ls01, yls99,gps14,dlz14}) 
and is analogous to the nucleon pairing models 
used to explain observations of cooling isolated NSs (\citealt*{plps04,gkyg04,gkyg05,syhhp11,plps13}).

With decrease of the stellar temperature $T$ 
the size of SFL-region 
[the region where neutrons are superfluid, it is given by the condition 
$T<T_{\rm cn}(x)$] 
either increases or, at sufficiently low $T$, 
remains unchanged. 
For instance, 
SFL-region corresponding to $T = 4\times10^8$~K, 
is shaded in Fig.\ \ref{Fig:Tc}.
One can see that the three-layer configurations are possible 
in some temperature range,
with no neutron superfluidity in the centre and in the
outer region, but with the superfluid intermediate region. 
At lower temperatures the star becomes a two-layer one, 
with the inner superfluid and outer normal region.

\begin{figure}
    \begin{center}
        \leavevmode
        \epsfxsize=7in \epsfbox{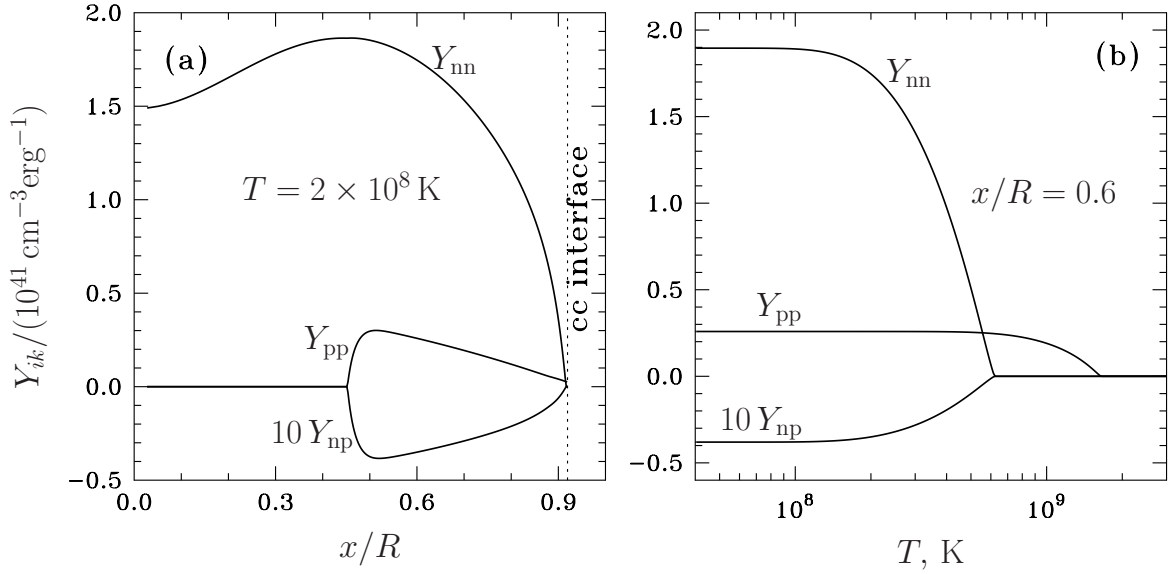}
    \end{center}
    \caption{Elements of the entrainment matrix $Y_{ik}$ 
    	(in units of $10^{41}\,\rm cm^{-3}\, erg^{-1}$) versus $x/R$ at $T=2\times 10^8 \,\rm K$ (panel a) 
    and versus temperature $T$ at $x/R=0.6$ (panel b).
		}
    \label{Fig:Yik}
\end{figure}

\begin{figure}
    \begin{center}
        \leavevmode
        \epsfxsize=7in \epsfbox{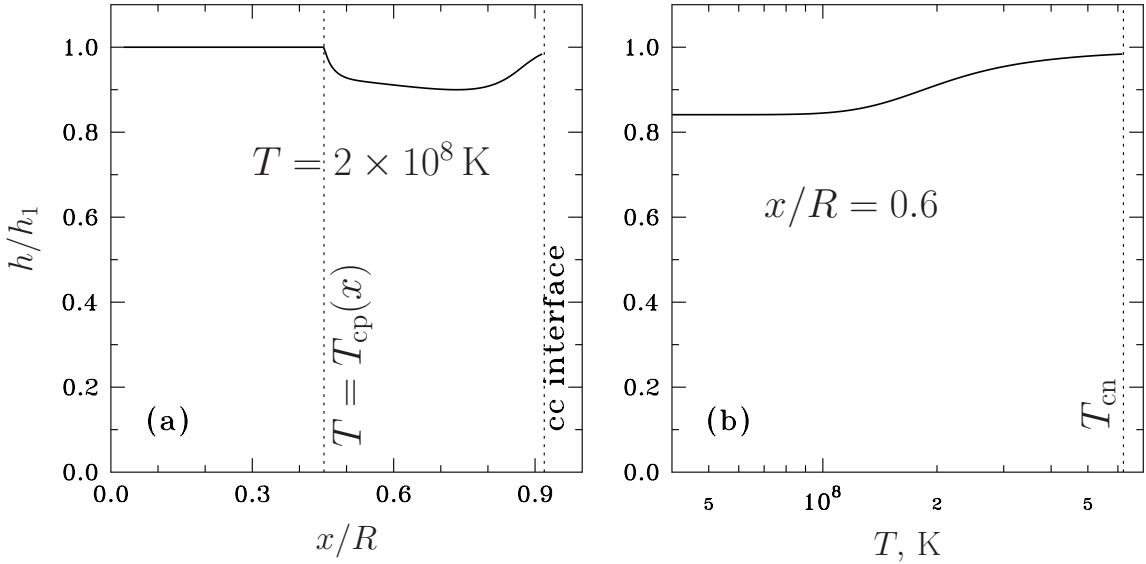}
    \end{center}
    \caption{The ratio $h/h_1$ versus $x/R$ at $T=2\times 10^8 \,\rm K$ (panel a) and versus temperature $T$ at $x/R=0.6$ (panel b).
		}
    \label{Fig:hh1}
\end{figure}

\begin{figure}
    \begin{center}
        \leavevmode
        \epsfxsize=7in \epsfbox{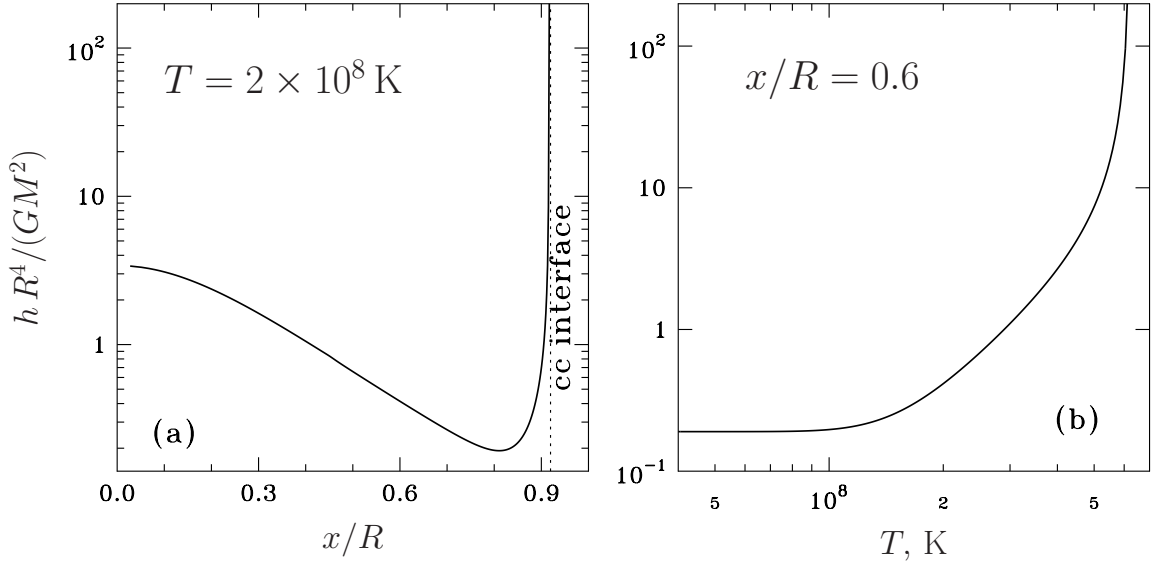}
    \end{center}
    \caption{Function $h$ (in units of $GM^2/R^4$) versus $x/R$ at $T=2\times 10^8 \,\rm K$ (panel a) and versus temperature $T$ at $x/R=0.6$ (panel b).
		}
    \label{Fig:beta}
\end{figure}

The symmetric entrainment matrix $Y_{ik}$
that parametrizes the effects of superfluidity in the oscillation equations,
is calculated in a way similar to how it was done in \cite{kg11}. 
Its elements are presented in Fig.\ \ref{Fig:Yik} 
as functions of the normalized radial coordinate $x/R$
at fixed temperature $T=2\times 10^8\,\rm K$ (panel a), 
and as functions of $T$ at fixed  $x/R=0.6$ (panel b).
One can see that the non-diagonal element $Y_{\rm np}$ 
is much smaller than the diagonal ones 
($Y_{\rm np}\sim 0.1 Y_{\rm pp}$ 
and $Y_{\rm np}\sim 0.02 Y_{\rm nn}$) 
and, in addition, 
$Y_{\rm np}$ vanishes when protons are non-superfluid. 
More realistic mean-field models, 
(see, e.g., \citealt{gkh09a,gkh09b,ghk14}), 
give similar or slightly higher 
ratios 
(by a factor of $2-3$ in the cited papers). 
This motivates us to consider the limit $Y_{\rm np}=0$ in this paper. 
In this limit $h(x)=h_1(x)$ and superfluid equation (\ref{sfl1}) 
reduces
to (\ref{sfl}), 
see section \ref{equations}. 
To further justify the assumption $Y_{\rm np}=0$, 
we plot the ratio $h/h_1$ in Fig. \ref{Fig:hh1}, 
which illustrates that 
$h(x)$ indeed approximately coincides with $h_1(x)$.
Panels (a) and (b) of Fig. \ref{Fig:hh1} show $h/h_1$ 
as a function of 
$x/R$
at $T=2\times 10^8\,\rm K$ 
and as a function of $T$ at 
$x/R=0.6$, 
respectively. 
To plot the figure, 
we used equations (\ref{beta}), (\ref{gamma}) and (\ref{y}) 
with non-zero $Y_{\rm np}$.
The parameter $h$, entering the superfluid equation (\ref{sfl}), 
is plotted in Fig.\ \ref{Fig:beta} 
as a function of $x/R$
at $T=2\times 10^8\,\rm K$ (panel a) 
and as a function of $T$ at 
fixed $x/R=0.6$ (panel b). 
To calculate it we used equations\ (\ref{beta}) and (\ref{y}) with $Y_{\rm np}=0$.

\section{Results for the spectrum}
\label{res}

\subsection{Model I, no muons}
\label{resI}

\begin{figure}
    \begin{center}
        \leavevmode
        \epsfxsize=7in \epsfbox{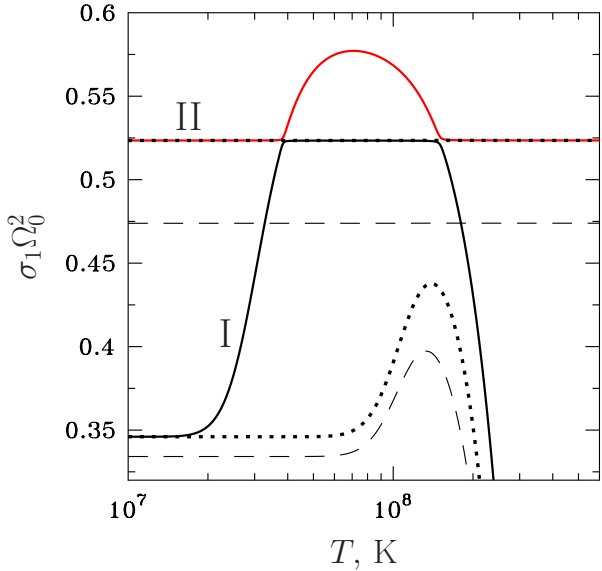}
    \end{center}
    \caption{(color online) $\sigma_1$ versus stellar temperature $T$. 
    Solid lines represent the spectrum calculated for $M=1.4M_\odot$ 
    and the critical temperature profiles shown in Fig. \ref{Fig:Tc}; 
    dots and dashes correspond to the density independent critical temperatures 
    ($T_{\rm cn}=6 \times 10^8\,\rm K$, 
    $T_{\rm cp}=5 \times 10^9\,\rm K$) 
    and stellar masses $M=1.4M_\odot$ and $M=1.7M_\odot$, respectively.
		}
    \label{Fig:spectrumnpe}
\end{figure}

\begin{figure}
    \begin{center}
        \leavevmode
        \epsfxsize=5in \epsfbox{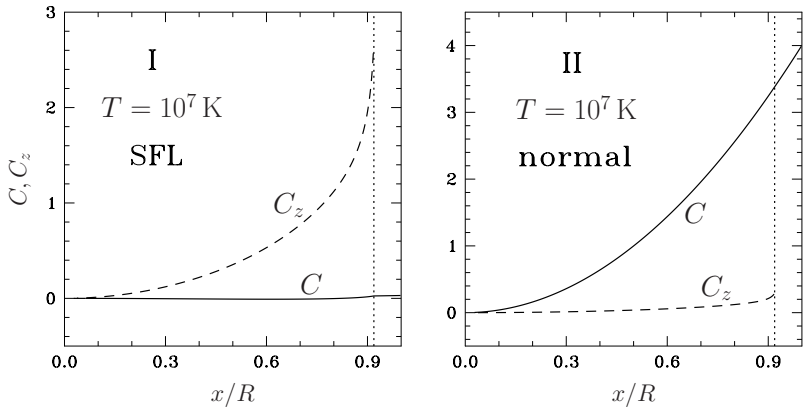}
    \end{center}
    \caption{Eigenfunctions $C$ (solid lines) and $C_z$ (dashed lines) 
    versus normalized radial coordinate $x/R$ calculated at $T=10^7\,\rm K$ for the mode I 
    (superfluid at this temperature, left panel) and the mode II (normal at this temperature, right panel). 
    Vertical dots show the crust-core interface. 
    Neutron superfluidity spans from the stellar centre to the crust-core interface at $T=10^7\,\rm K$.
		}
    \label{Fig:efnpe}
\end{figure}

Below we shall present the numerical results only for the modes with $m=2$.
These modes are the most interesting for us, 
since they can interact resonantly (see section \ref{discus})
with the normal r-mode with $m=2$,
which is known to be the most unstable inertial mode in NSs.
The results of our numerical solution 
to the system of differential equations (\ref{system}) 
with the boundary conditions from section \ref{bc} 
are presented in Fig. \ref{Fig:spectrumnpe}, 
where the function $\sigma_1(T)$ 
is shown by solid lines.
It is calculated by employing
the physics input 
from section \ref{PhysInput} 
(we remind that $\sigma_0$ 
is the same for all the toroidal modes with $\sigma \propto \Omega$).
We find that a superfluid NS with the simplest $\rm npe$-composition of the core 
harbours only two toroidal modes with $\sigma \propto \Omega$ 
(marked I and II in Fig. \ref{Fig:spectrumnpe}). 
At any given temperature (except near avoided crossings)
these two modes, 
are of clearly distinct nature, either normal or superfluid.
Normal mode is 
practically independent of temperature and 
almost coincides with 
the ordinary r-mode in a normal (non-superfluid) star.
In contrast, superfluid mode depends on temperature 
via
(i) temperature-dependent function $h$ 
and 
(ii) shrinking of the superfluid region with increasing $T$.
Figure \ref{Fig:efnpe} presents the eigenfunctions 
for these two modes
versus radial coordinate $x$ 
in the low-temperature limit.
One can clearly see that the normal mode corresponds 
to co-motion, 
while the superfluid mode --- to counter-motion 
of superfluid and normal liquid components.
However, the situation is not so distinct 
in the vicinity of the avoided-crossings 
($T\approx 3.8\times 10^7\,\rm K$ and 
$T\approx 1.5\times 10^8\,\rm K$ in our numerical example), 
where the 
modes change their behaviour
from normal-like to superfluid-like and vice-versa. 
Near these `resonance' temperatures 
the modes are `hybrids' with the mixed properties
(see \citealt{cg11,gkcg13,gkgc14,gck14a,gck14b} for a detailed discussion).

To illustrate the sensitivity of the oscillation spectrum 
to the model of superfluidity 
and the stellar mass, 
we carried out 
the same calculations for the density-independent 
critical temperatures 
$T_{\rm cn}=6 \times 10^8\,\rm K$ and
$T_{\rm cp}=5 \times 10^9\,\rm K$, 
assuming two different stellar masses:
(i) $M=1.4M_\odot$ 
(the results are shown by dots in Fig.\ \ref{Fig:spectrumnpe}), 
and (ii) $M=1.7M_\odot$ 
(dashes in Fig.\ \ref{Fig:spectrumnpe}). 
One can see that the superfluid mode 
is sensitive to the critical temperature profiles, 
while the normal mode is not. 
At the same time, the variation of the stellar mass affects both modes.

\subsection{Model II, muons are present}
\label{resII}

\begin{figure}
    \begin{center}
        \leavevmode
        \epsfxsize=7in \epsfbox{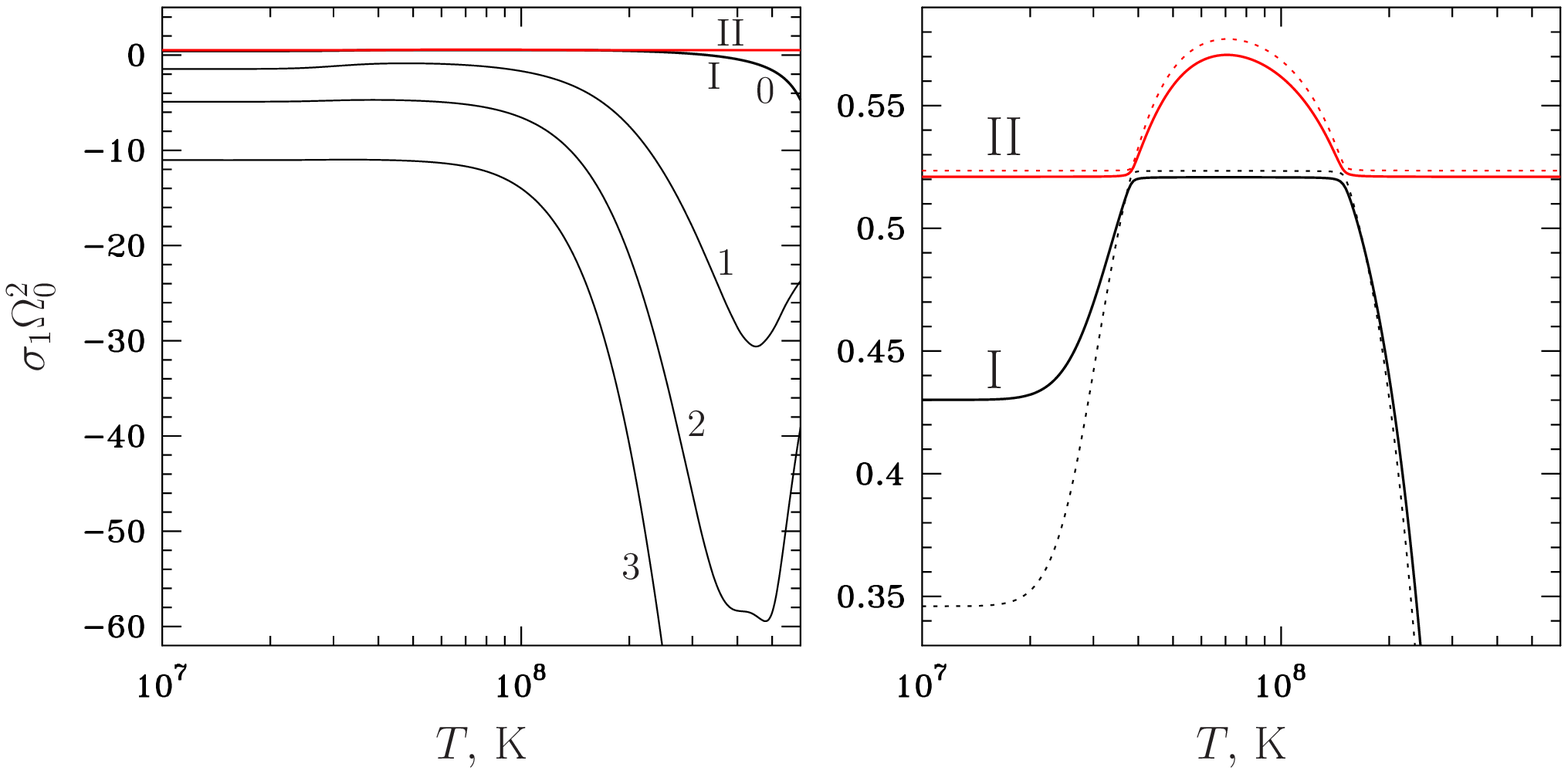}
    \end{center}
    \caption{(color online) The fragment of the spectrum 
    of a $1.4 M_\odot$ superfluid NS with $\rm npe\mu$-composition of the core.
    Left panel shows the function $\sigma_1(T)$ for 
    two nodeless r-modes and a set of superfluid r-modes, 
    differing by the number of nodes,
    $n=1$, $2$, and $3$ (number of nodes, including $n=0$, 
    is indicated near the curves).
    The right panel is a 
    zoomed in version of the left one, 
    it contains two nodeless r-modes. 
    Dots in the right panel represent
    the spectrum calculated for $\rm npe$ composition of the core
    and the same stellar mass $M=1.4M_\odot$.
		}
    \label{Fig:spectrumnpemu}
\end{figure}

The numerical results for the spectrum 
of a NS containing muons in the core 
are shown in two panels of Fig. \ref{Fig:spectrumnpemu}. 
The right panel is 
a strongly 
zoomed in version
of the left one. 
We find that an admixture of muons, 
populates
the NS spectrum of toroidal modes
(in addition to two nodeless r-modes, 
	existing
	in $\rm npe$ NSs)
with an infinite set of superfluid modes, 
which differ from one another 
by the number of nodes $n$ of the superfluid function $C_z(x)$.
We show only  the first four superfluid modes with $n=0$, $1$, $2$, $3$ in the left panel of the figure. 
The right panel shows the fragment of the left one which contains only two nodeless r-modes.

\begin{figure}
    \begin{center}
        \leavevmode
        \epsfxsize=5in \epsfbox{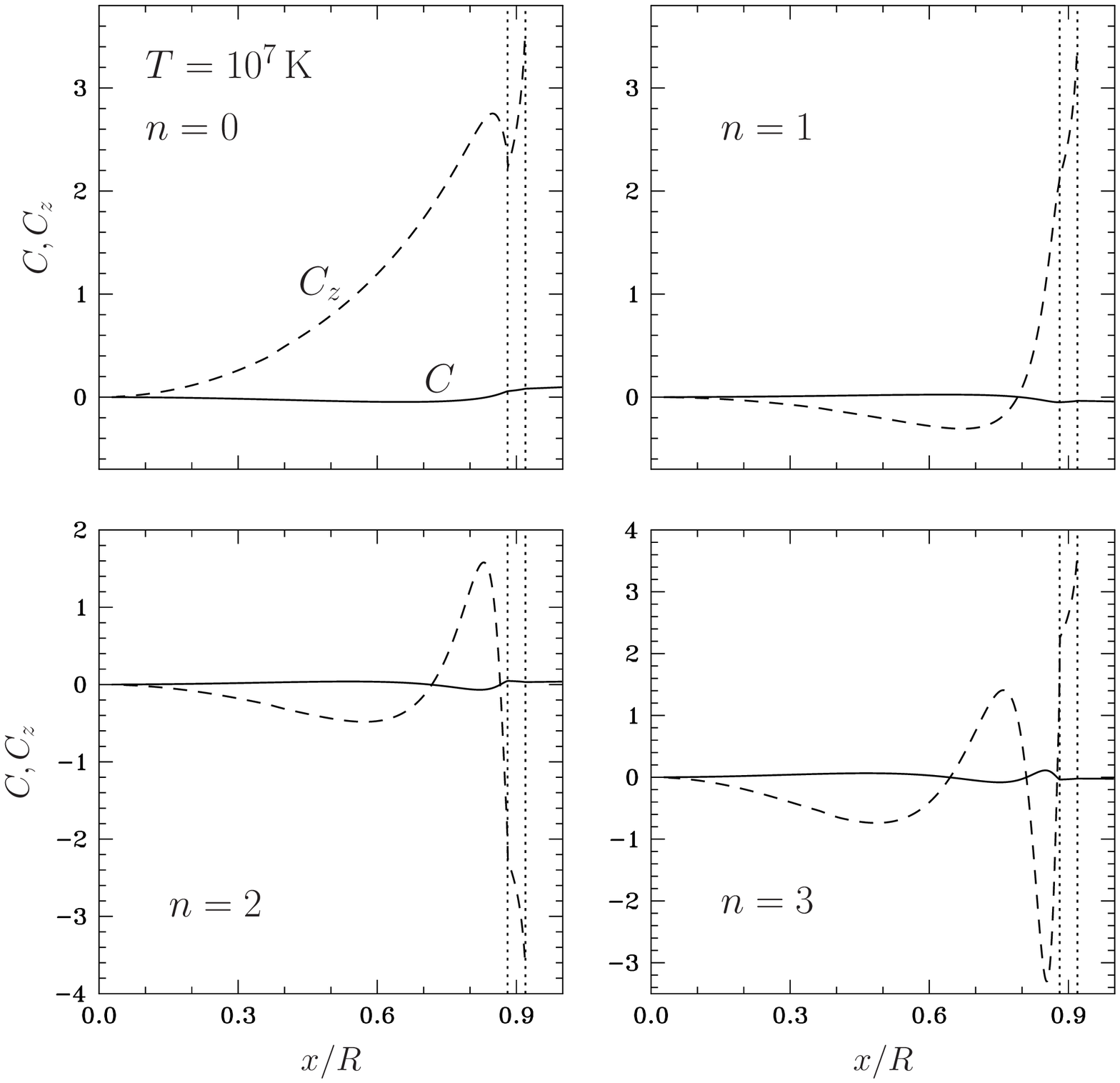}
    \end{center}
    \caption{Eigenfunctions $C$ (solid lines) 
    and $C_z$ (dashed lines)  versus $x/R$ 
    for superfluid modes with the number of nodes $n=0$, $1$, $2$, and $3$.
    The figure is plotted for $T=10^7\,\rm K$, 
    at which the entire core is occupied by the neutron superfluidity.
    Vertical dots show (from left to right) the muon threshold density 
    	and the crust-core interface. 
		}
    \label{Fig:efnpemu}
\end{figure}

\begin{figure}
    \begin{center}
        \leavevmode
        \epsfxsize=3in \epsfbox{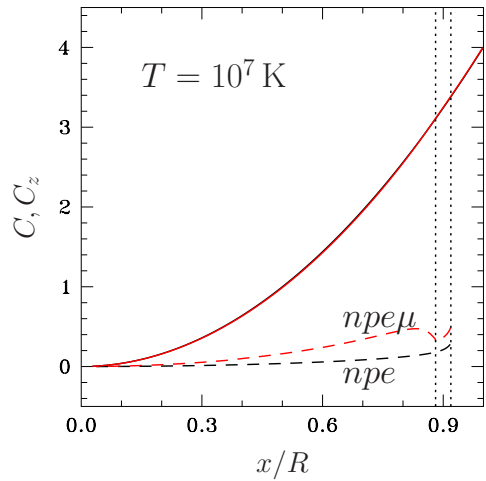}
    \end{center}
    \caption{(color online) Eigenfunctions $C$ (solid lines) 
    and $C_z$ (dashed lines) for normal r-mode versus $x/R$ 
    calculated at $T=10^7\,\rm K$ for $\rm npe$ and $\rm npe\mu$ core compositions. 
    Vertical dots show (from left to right) the muon threshold density and the crust-core interface. 
		}
    \label{Fig:efcompare}
\end{figure}

The eigenfunctions for the first four superfluid modes 
(one nodeless mode and three modes with nodes) 
are plotted in Fig.\ \ref{Fig:efnpemu} 
assuming $T=0$.
Each panel is marked with the number of nodes 
of the superfluid function $C_z(x)$.
%
%
%
Although the muon presence results in additional 
superfluid toroidal modes with $n\neq0$
in the spectrum, 
they do not affect 
much 
the eigenfrequencies of the modes without nodes. 
To illustrate this point, 
we show by dots the spectrum of an NS with the npe-composition of the core
in the right panel of Fig.\ \ref{Fig:spectrumnpemu}. 
Notably, the difference 
is stronger at low temperatures, 
at which the Brunt-V$\ddot{\rm a}$is$\ddot{\rm a}$l$\ddot{\rm a}$ frequency 
is higher and stratification is more pronounced (\citealt{kg14}). 
Figure \ref{Fig:efcompare} illustrates how the admixture of muons affects 
the eigenfunctions of normal r-mode. 
Solid lines and dashed lines show, 
respectively, the functions $C(x)$ and $C_z(x)$ 
for $\rm npe$ and $\rm npe\mu$ core compositions. 
$C(x)$ appears to be almost insensitive to the composition 
(two lines are practically indistinguishable in the figure), 
while the superfluid displacement $C_z(x)$ 
is substantially larger in the  
$\rm npe\mu$ case.
Thus, one can expect that the core composition 
may affect the r-mode dissipation due to mutual friction, 
which is sensitive to the difference between the normal and superfluid velocities (\citealt{mendell91,lm00,asc06}),
that is to the function $C_z(x)$ (see section \ref{damp}).

\section{Why the results for $\rm npe$ and $\rm npe\mu$ core compositions are so different?}
\label{diff}

Our numerical 
analysis
reveal that 
NSs with $\rm npe$ core composition 
host
only two nodeless r-modes 
(i.e., normal and superfluid r-modes), 
while account for muons leads, in addition, to
an infinite set of superfluid r-modes with the nodes. 
What is the reason for such a difference?
To answer this question, 
consider the continuity equations for leptons. 
In the coordinates $r$, $\theta$, $\phi$ they read:
\begin{eqnarray}
\delta n_{\rm l}+\frac{1}{r^2}\frac{\partial}{\partial r}\left(n_{\rm l} \xi_r r^2\right)+\frac{1}{r{\rm sin}\theta}\frac{\partial}{\partial \theta}\left(n_{\rm l} \xi_\theta {\rm sin}\theta\right)+\frac{\imath m}{r{\rm sin}\theta}n_{\rm l} \xi_\phi=0.
\end{eqnarray}
For inertial modes $\delta n_{\rm l}$ vanishes to the leading order in $\Omega/\Omega_0$ 
(i.e., $\delta n_{\rm l}=O(\Omega^2/\Omega_0^2)$, see equation \ref{deltaf}), 
while $n_{\rm l}$ can be considered as 
$\theta$-independent 
and factored out of the corresponding derivative,
\begin{eqnarray}
\frac{1}{r^2}\frac{\partial}{\partial r}\left(\xi_r r^2\right)+\frac{1}{r{\rm sin}\theta}\frac{\partial}{\partial \theta}\left(\xi_\theta {\rm sin}\theta\right)+\frac{\imath m}{r{\rm sin}\theta}\xi_\phi+\frac{\xi_r}{n_{\rm l}}\frac{\partial n_{\rm l}}{\partial r}=0.
\end{eqnarray}
Combining equations for electrons and muons, we find
\begin{eqnarray}
\frac{\xi_r}{n_{\rm e}}\frac{\partial n_{\rm e}}{\partial r}=\frac{\xi_r}{n_{\rm \mu}}\frac{\partial n_{\rm \mu}}{\partial r}.
\end{eqnarray}
Since, generally, 
$\partial (n_{\rm e}/n_{\rm \mu})/\partial r\neq 0$, 
$\xi_r$ has to vanish 
in the leading order in $\Omega/\Omega_0$,
$\xi_r=O(\Omega^2/\Omega_0^2)$, for inertial modes
in the superfluid $\rm npe\mu$ matter. 
This conclusion, of course, does not 
concern
the functions
$\xi_{{\rm b}r}$ and $z_r$. 
In the superfluid $\rm npe$ matter 
such 
constraint
is absent,
because superfluid neutrons do not move with the same velocity as electrons, 
while protons are non-stratified ($n_{\rm p}=n_{\rm e}$).  

To put it differently, superfluidity makes $\rm npe$ matter non-stratified
(the corresponding Brunt-V$\ddot{\rm a}$is$\ddot{\rm a}$l$\ddot{\rm a}$ 
frequency vanishes 
if we neglect small entropy contribution, see \citealt{gk13}), 
allowing for non-zero $\xi_r$, 
while admixture of muons 
again stratifies it (\citealt{kg14,dg16,pah16}), 
requiring $\xi_r=0$ at $\Omega \rightarrow 0$. 
The situation is very similar to what happens 
in non-superfluid NSs: 
In a stratified star $\xi_r$ 
must vanish at $\Omega \rightarrow 0$ (\citealt{yl00}), 
but
this is not necessary in non-stratified stars.

As in non-superfluid NSs, 
the condition $\xi_r=0$ transforms 
superfluid inertial modes (mixed modes with non-zero $\xi_r$)
in $\rm npe$-matter 
to purely toroidal r-modes in $\rm npe\mu$-matter, 
producing thus an infinite set of superfluid r-modes which differ by the number of nodes $n$
of the eigenfunction $C_z(x)$
(see section \ref{discus} for more details).

\section{Dissipative damping of r-modes}
\label{damp}

Let us now illustrate how the avoided-crossings of r-modes 
affect their dissipative damping. 
At resonance temperatures 
(near avoided-crossings) the most effective dissipation mechanism 
is the mutual friction (\citealt{als84,lm00,ly03,gck14a,gck14b}). 
According to \cite{als84}, it arises from the electron scattering 
off the neutron vortices magnetized 
due to 
entrainment between the superfluid neutrons and superconducting protons. 
Note that, this mechanism is relevant only when $Y_{\rm np}\neq 0$. 
Thus, although we set $Y_{\rm np}= 0$ when calculating the r-mode
eigenfunctions and eigenfrequencies 
(which should be a reasonable approximation for sufficiently small $Y_{\rm np}$, 
see a corresponding discussion in section \ref{discus}),
we, at the same time, assume that $Y_{\rm np}$ is finite when studying the r-mode damping.

The general relativistic expression for the oscillation energy dissipation rate due to mutual friction
(per unit volume), $\dot{\epsilon}_{\rm MF}$,
can be deduced from \cite{gd16} (see also \citealt{kg12,dg16,gusakov16}). 
In the Newtonian limit, we are interested in here,
the corresponding expression can be obtained by making use of the 
equation (90) from \cite{gd16}, 
as well as the
formulas (\ref{sfl_gen}) and (\ref{f_gen}) and the definition of ${\pmb z}$.
As a result, we arrive at the following general formula (in dimensional units),
\begin{eqnarray}
\dot{\epsilon}_{\rm MF}=\frac{2\beta_{\rm MF}}{\Omega}\sigma^2 n_{\rm b}^2\frac{ \mu_{\rm n}^2}{c^4}\left[(\bf{\Omega z})^2-\bf{z}^2 \bf{\Omega}^2\right], \label{EMF}
\end{eqnarray}
which reduces to (up to the leading order in the rotation frequency)
\begin{eqnarray}
\dot{\epsilon}_{\rm MF}=-2\beta_{\rm MF}\Omega\sigma^2 n_{\rm b}^2\frac{ \mu_{\rm n}^2}{c^4}\left(z_\theta^2{\rm cos}^2\theta+z_\phi^2\right) \label{epsilondot}
\end{eqnarray}
in the case of r-modes.
Here $\beta_{\rm MF}$ is the mutual friction coefficient 
from the formula (\ref{f_gen}) (the same coefficient is
introduced
in \citealt{khalatnikov89,ss95,kg12,gusakov16,dg16,gd16}). 
In the NS literature the dimensionless coefficient $B$ is often used 
instead of $\beta_{\rm MF}$ 
(see, e.g., \citealt{mendell91,asc06,hap09}; 
note that the coefficient $B$ in \citealt{mendell91,asc06,hap09} differs from $B$ in \citealt{hv56}!). 
In the zero temperature limit they are related by 
$B=\beta_{\rm MF} \,n_{\rm n} \mu_{\rm n}/c^2$ 
%
\footnote{To obtain this relation, 
compare the equation on the neutron superfluid velocity (I7) from \cite{gd16} 
with the mutual friction force given by (I8) and, 
e.g., corresponding equations (2) and (3) in \cite{hap09}. 
Note that the vector $\bf{w}_{\rm np}$ in \cite{hap09} equals the vector $\bf{W}$ in \cite{gd16}.}.
%
Using this relation one can check that 
equation (\ref{EMF}) reduces to the equation (114) of \cite{hap09}.

\subsection{Mutual friction coefficient}
\label{mfcoef}

To calculate the mutual friction coefficient $\beta_{\rm MF}$ 
one needs to know the time-scale $\tau_{\rm v}$ on which electrons 
(and other species strongly coupled 
to
 electrons, i.e.
muons, protons, and normal neutron excitations; hereafter the normal liquid component) 
relax to the motion of the neutron vortices. 
For cold $\rm npe$-matter it was derived in \cite{als84} (their equation 30b). 
At a finite temperature 
and in $\rm npe\mu$ matter $\tau_{\rm v}$ has not been estimated yet. 
In that case not only electrons, but also muons and proton thermal excitations
can scatter off the magnetized neutron vortices.
Moreover, at finite temperatures both neutron and proton thermal excitations 
can, in principle, scatter off the neutron excitations localized in the vortex cores. 
These issues clearly deserve a separate study. 
Here, for definiteness, we shall use the result of \cite{als84} 
for cold $\rm npe$-matter, only slightly modifying it. 
Namely, we shall respect that electrons are coupled not only to protons 
but also to muons and normal neutron excitations.
Then the relaxation time can be presented as (\citealt{als84})
\begin{eqnarray}
\tau_{\rm v}^{-1}=3\,
\frac{\mu_{\rm e} n_{\rm e}}{\mu_{\rm n} n_{\rm b}-\mu_{\rm n} \mu_k Y_{{\rm n}k}}
\tau_0^{-1}\alpha^{-3}\beta^4 \int^\alpha_0 \frac{x^2+\alpha^2}{\left(x^2+\beta^2\right)^2} \left|\frac{J_1(x)}{x}\right|^2 dx
\equiv \frac{3\pi}{16}\frac{\mu_{\rm e}}{\mu_{\rm p}} \tau_0^{-1}\frac{\beta}{\alpha}\left[1-g(\beta)\right], 
\label{tauv}
\end{eqnarray}
where $\mu_{\rm n} n_{\rm b}-\mu_{\rm n} \mu_k Y_{{\rm n}k}$ 
represents the density of the normal liquid component (\citealt{ga06}), 
and the last equality should be considered as the definition of the function $g(\beta)$. 
In equation (\ref{tauv}) $J_1$ is the Bessel function of the first kind; 
$\alpha=2k_{\rm e} \xi$, $\beta=\xi/\Lambda_\star$, $\tau_0^{-1}=\pi N_\tau \Phi_\star^2$, 
where $\xi$ is the neutron coherence length, 
$\Lambda_\star=\left(4\pi e^2 Y_{\rm pp}\right)^{-1/2}$ is the London penetration depth, 
$k_{\rm e}$ is the electron Fermi wave vector, 
$\Phi_\star=\frac{\pi \hbar c}{e}\frac{Y_{\rm np}}{Y_{\rm pp}}$ is the flux of the neutron vortex line, 
and, finally,
\begin{eqnarray}
N_\tau=\frac{2\pi}{\hbar}n_{\rm v}\left(\frac{e\hbar}{2m_{\rm e} c}\right)^2\left(\frac{m_{\rm e} c^2}{\mu_{\rm e}}\right)^2\frac{\mu_{\rm e}}{(\pi\hbar c)^2},
\end{eqnarray}
where
$n_{\rm v}$ 
is 
the vortex density.
It is possible to express (see appendix \ref{mf}) $\beta_{\rm MF}$ through $\tau_{\rm v}$
	by the formula
\begin{eqnarray}
\beta_{\rm MF}=
\frac{\mu_{\rm n} n_{\rm b}-\mu_{\rm n} \mu_k Y_{{\rm n}k}}{(\mu_{\rm n} \mu_k Y_{{\rm n}k})^2}\frac{c^2}{2\Omega \tau_{\rm v}}.
\end{eqnarray}

\begin{figure}
    \begin{center}
        \leavevmode
        \epsfxsize=3in \epsfbox{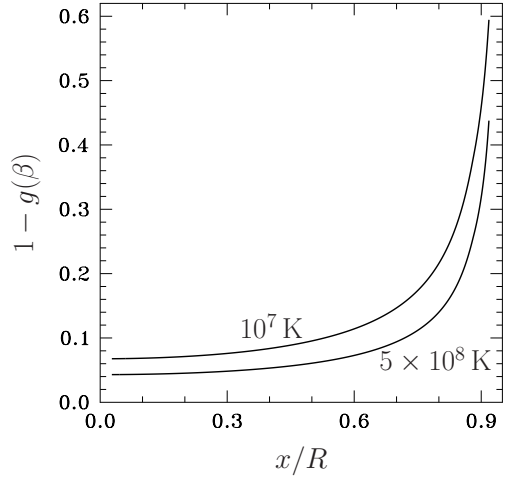}
    \end{center}
    \caption{The function $1-g(\beta)$ (see equation \ref{tauv}) versus $x/R$ calculated at $T=10^7\,\rm K$ (upper curve) and $T=5\times 10^8\,\rm K$ (lower curve); $T_{\rm cn}=6\times 10^8\,\rm K$, $T_{\rm cp}=5\times 10^9\,\rm K$.
		}
    \label{Fig:g}
\end{figure}

Both the coefficient $\beta_{\rm MF}$ (or $B$) and the relaxation time $\tau_{\rm v}$ 
depend on the entrainment matrix elements $Y_{ik}$, 
which are, generally, functions of the Landau parameters $f_1^{ik}$  and temperature (\citealt{gh05,gkh09a,gkh09b}).
We would also like to note that an approximate formula for $B$, 
adopted in the literature 
(see, e.g., equation 66 in \citealt{asc06}), 
can hardly be a good approximation even in the zero-temperature limit, since it assumes $g(\beta)=0$. 
We plot
$1-g(\beta)$ as a function of $x/R$ in Fig.~\ref{Fig:g} to demonstrate that the coefficient $1-g(\beta)$ in equation (\ref{tauv}) 
is not close to 1.
The curves correspond to two stellar temperatures ($T=10^7\,\rm K$ and $T=5\times 10^8\,\rm K$);
the figure is plotted for constant
$T_{\rm cn}=6\times 10^8\,\rm K$ and $T_{\rm cp}=5\times 10^9\,\rm K$.
%

\subsection{Dissipation rate and energy of oscillations }
\label{integrals}

The mutual friction damping time for r-modes 
can be defined as (\citealt{lm00,ly03,hap09,agh09})
\begin{eqnarray}
\tau_{\rm MF}=-\frac{2E}{\dot{E}_{\rm MF}}, 
\label{tau}
\end{eqnarray}
where $E$ is the mechanical energy of oscillations
coinciding, 
to the leading order in $\Omega/\Omega_0$, 
with their
kinetic energy 
(e.g., \citealt{lm00,hap09}). 
The latter is given by (\citealt{gkcg13})
\begin{eqnarray}
E=\int \epsilon_{\rm kin}dV=\int \frac{1}{2}\sigma^2 \frac{n_{\rm b}\mu_{\rm n}}{c^2}\left(\xi_{\rm b}^2+y z^2\right) dV.
\end{eqnarray}
Substituting the real parts of $\xi_{\rm b}$ and $z$ from (\ref{xib0}) and (\ref{z0}), respectively, we get
\begin{eqnarray}
E=\frac{\sigma^2}{2 c^2}\int_0^R n_{\rm b}(x)\mu_{\rm n}(x) \left[C^2(x)+y(x) C_z^2(x)\right] x^2 dx \int_0^{2\pi}d\phi \nonumber \\
\int_0^{\pi}  {\rm sin}\theta \,d\theta
\left[m^2 P_m^m({\rm cos}\theta)^2 {\rm sin}^{-2}\theta\,{\rm sin}^2(\sigma t+m\phi)+
\left(\frac{d P_m^m({\rm cos}\theta)}{d\theta}\right)^2{\rm cos}^2(\sigma t+m\phi)\right].
\end{eqnarray}
Integration of this formula over $\phi$ makes the energy independent of 
the phase of oscillations 
and introduces an additional factor 
$\pi$, 
while integration over $\theta$ results 
in a factor $2m(m+1)(2m)!/(2m+1)$, so that 
\begin{eqnarray}
E=\frac{\Omega^2\sigma_0^2}{c^2}\pi\frac{m(m+1)(2m)!}{2m+1}\int_0^R n_{\rm b}(x)\mu_{\rm n}(x) \left[C^2(x)+y(x) C_z^2(x)\right] x^2 dx.
\end{eqnarray}

Next, $\dot{E}_{\rm MF}$ in equation (\ref{tau}) 
is the rate of oscillation energy damping due to mutual friction,
\begin{eqnarray}
\dot{E}_{\rm MF}=\int\dot{\epsilon}_{\rm MF}dV.
\end{eqnarray}
(Actually, the integrand is non-zero only 
in the region where both neutrons and protons are superfluid.)
Substituting (\ref{epsilondot}) with 
$z_\theta$ and $z_\phi$ from (\ref{z0}), we obtain
\begin{eqnarray}
\dot{E}_{\rm MF}=-\frac{2\Omega\sigma^2}{c^4} \int_0^R 
\beta_{\rm MF}(x)
n_{\rm b}^2(x) \mu_{\rm n}^2(x)C_z^2(x) x^2 dx\int_0^{2\pi}d\phi \nonumber\\
\int_0^{\pi} 
\left[m^2 P_m^m({\rm cos}\theta)^2 {\rm tan}^{-2}\theta\,{\rm sin}^2(\sigma t+m\phi)+
\left(\frac{d P_m^m({\rm cos}\theta)}{d\theta}\right)^2{\rm cos}^2(\sigma t+m\phi)\right]  {\rm sin}\theta \,d\theta.
\end{eqnarray}
Integration over $\theta$ and $\phi$ gives 
\begin{eqnarray}
\dot{E}_{\rm MF}=-\frac{\Omega^3\sigma_0^2}{c^4}\pi\frac{4m(2m)!}{2m+1} \int_0^R \beta_{\rm MF}(x) n_{\rm b}^2(x) \mu_{\rm n}^2(x)C_z^2(x) x^2 dx,
\end{eqnarray}
which is independent of the phase of oscillations.
 Then
\begin{eqnarray}
\tau_{\rm MF}=\frac{m+1}{2\Omega}c^2\frac{\int_0^R n_{\rm b}(x)\mu_{\rm n}(x) \left[C^2(x)+y(x) C_z^2(x)\right] x^2 dx}{\int_0^R \beta_{\rm MF}(x) n_{\rm b}^2(x)  \mu_{\rm n}^2(x)C_z^2(x) x^2 dx}\equiv \tau_{0\,\rm MF}\frac{\Omega_0}{\Omega}, \label{tau1}
\end{eqnarray}
where the last equality is the definition of $\tau_{0\,\rm MF}$
\footnote{Note that both $C(x)$ and $C_z(x)$ excite in the leading order in $\Omega$, 
thus the ratio of integrals in equation (\ref{tau1}) is independent of the rotation frequency.}.

\subsection{Results for damping times and instability window}
\label{resdamp}

\begin{figure}
    \begin{center}
        \leavevmode
        \epsfxsize=3in \epsfbox{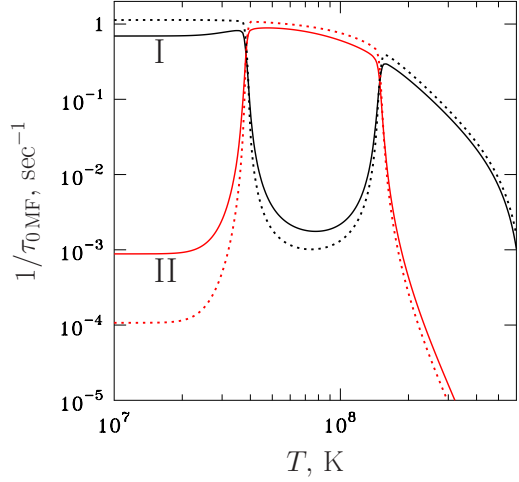}
    \end{center}
    \caption{Inverse damping time $\tau_{0\,\rm MF}$ (see equation \ref{tau1}) 
    	versus $T$ for the two nodeless r-modes from section \ref{res}
    	(solid lines for $\rm npe\mu$ matter, dots for $\rm npe$ matter). 
		}
    \label{Fig:tau}
\end{figure}

Figure~\ref{Fig:tau} shows the inverse damping time, $\tau_{0\,\rm MF}$, 
(see equation \ref{tau1}) as a function of temperature $T$ 
for two nodeless r-modes (I and II, see Figs.\ \ref{Fig:spectrumnpe} and \ref{Fig:spectrumnpemu}). 
$C_z(x)$ is larger
for the superfluid mode, 
hence it dissipates
stronger, as it should be (see equation \ref{tau1}). 
Correspondingly, when a mode changes its behaviour from normal-like to superfluid-like, 
the efficiency of mutual friction dissipation for that mode rises sharply. 
The rapid decrease of mutual friction efficiency 
at high temperatures (at $T\ga 2\times 10^8$~K in the figure) 
is due to decrease of the entrainment matrix element $Y_{\rm np}$ 
at $T\rightarrow T_{\rm cn}$. 
Normal mode in $\rm npe\mu$ matter (solid lines) dissipates faster than in $\rm npe$ matter (dots) 
because $C_z(x)$ is higher in $\rm npe\mu$ matter (see Fig. \ref{Fig:efcompare}).

Having determined the mutual friction damping time $\tau_{\rm MF}$,
we can calculate 
the instability windows, 
that is 
the regions of stellar frequency and temperature 
at which a star becomes unstable with respect to excitation of r-modes.
They are determined by
the condition $1/\tau_{\rm GR}+1/\tau_{\rm Diss}<0$,
where $\tau_{\rm GR}$ is the gravitational radiation timescale (see, e.g., \citealt{ak01,gck14a})
and $\tau_{\rm Diss}$ is the dissipation timescale.
As discussed, e.g., in \cite{gck14a}, 
at the temperatures of interest the main dissipative mechanisms are 
the mutual friction dissipation 
and
shear viscosity.
Consequently,  
$1/\tau_{\rm Diss}=1/\tau_{\rm MF}+1/\tau_{\rm S}$, 
where $\tau_{\rm MF}$ is given by equation (\ref{tau1}) and
$\tau_{\rm S}$ (the dissipation timescale due to shear viscosity) is taken from \cite{gck14a}, 
see their formula (13)
\footnote{Formula (13) of \cite{gck14a} assumes a polytropic NS with strong proton superconductivity. 
Here we apply the same formula to our NS models with critical temperature profiles from Fig. \ref{Fig:Tc}.}.

\begin{figure}
    \begin{center}
        \leavevmode
        \epsfxsize= 3.0in \epsfbox{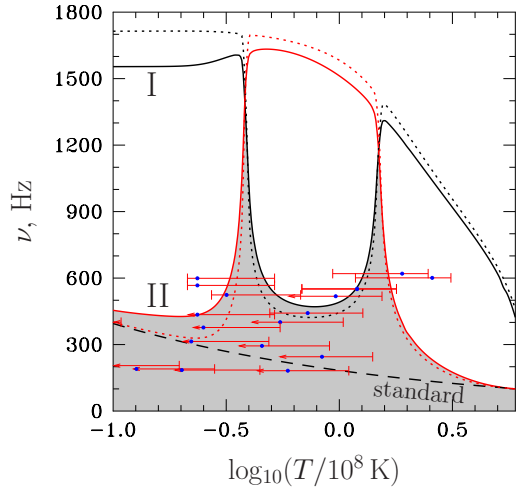}
    \end{center}
    \caption{Instability curves in the coordinates `stellar rotation frequency - internal temperature' for the two r-modes without nodes (I and II, solid lines for $\rm npe\mu$ composition, dots for $\rm npe$ composition). Dashes show the `standard instability curve', when normal r-mode is damped by shear viscosity only 
		(see \citealt{gck14a}). Circles with error bars represent observational data (\citealt{gck14a,cgk17}).}
    \label{Fig:nu}
\end{figure}

The instability windows, calculated as described above, 
are presented in Fig. \ref{Fig:nu} for two nodeless r-modes (I and II, see solid lines). 
In the region filled with grey 
the $\rm npe\mu$ NSs are stable (this is the `stability' region), 
in the white region 
they are
unstable. 
Circles with error bars show the observational data 
(taken from \citealt{gck14a,cgk17}). 
Dots 
indicate the instability window for $\rm npe$ NSs.

Generally, the stability region, calculated 
not accounting for the resonant interaction of oscillation modes 
(but under realistic assumptions about kinetic properties of NS matter)
is rather small and cannot contain the hottest 
and most rapidly rotating NSs 
(see, e.g., \citealt{hah11,hdh12,gck14a,cgk14,cgk17} 
and the dashed curve in Fig. \ref{Fig:nu}). 
Obviously, this fact poses a problem for the theory (\citealt{levin99,hah11,hdh12,gck14a,gck14b}).
On the other hand, the realistic stability region 
easily accommodates all the observed sources,
but it is obtained under a simplifying assumption $Y_{\rm np}=0$.
What would happen if we relaxed this assumption is discussed in the next section.

\section{Discussion}
\label{discus}

Prior to discussing the spectrum 
of SFL NSs let us first analyze the non-superfluid case. 
For that purpose the work of \cite{yl00} 
can be very helpful (see also \citealt{lf99}).
At high rotation frequencies the spectra of stratified 
and non-stratified NSs are similar 
--- there is one nodeless r-mode 
and an infinite set of inertial modes. 
At low rotation frequencies stratification transforms 
inertial modes (the modes with mixed poloidal and toroidal components) into either g-modes 
or r-modes with nodes and $l=m$. 
In accordance with the connection rule (21)--(22) 
of \cite{yl00}, 
inertial modes $i_k$ with $k\geq 1$ transform into g-modes, 
and $i_0$ modes transform into r-modes 
\footnote{\cite{yl00} 
expand the equations describing inertial modes 
into series in terms of the Legendre polynomials, $P_l^m({\rm cos}\theta)$.
For each $m$ there exists a system of independent equations
that couples
Legendre polynomials 
with different $l$. 
As a result, any inertial mode is a mixture 
of various Legendre polynomials $P_l^m({\rm cos}\theta)$ with fixed $m$.
It turns out that for any such mode one can identify
the leading toroidal harmonic
[$P_{l_0-1}^m({\rm cos}\theta)$ in notations of \citealt{yl00}],
so that 
it
can be characterized by  two angular
`quantum' numbers, $l_0$ and $m$.
However, after fixing $l_0$ and $m$ we still have some freedom
resulting in a
set of inertial modes with the frequencies $\sigma_k$.
\cite{yl00} enumerate these modes following the rule (for axial modes):
$0\leq \sigma_0 <\sigma_1 <...<\sigma_k < ... <\sigma_{(l_0-1-|m|)/2}$. 
Thus, generally, an inertial mode can be characterized by three numbers,  $l_0$, $m$, and $k$.
\label{foot_k}}.
This transformation occurs at rotation frequencies 
of the order of g-mode frequencies 
in a non-rotating NS. 
Thus, at slow rotation the toroidal modes 
with $\sigma \sim \Omega$ 
and $l=m$ of a \underline{stratified} star are: 
(i) one nodeless r-mode 
and 
(ii) an infinite set of r-modes with nodes. 
In \underline{non-stratified} stars there is only one nodeless r-mode.

The spectrum of SFL NSs is rather similar. 
In superfluid npe-matter 
the Brunt-V$\ddot{\rm a}$is$\ddot{\rm a}$l$\ddot{\rm a}$ frequency
vanishes,  
making NSs with $\rm npe$ core composition (hereafter $\rm npe$ NSs) non-stratified (\citealt{gk13}),
while even small admixture of muons 
stratifies them again (\citealt{kg14,dg16,pah16}).
Our calculations show that,
at low rotation rates (and at $Y_{\rm np}=0$) 
the only toroidal modes with $\sigma \sim \Omega$ 
in $\rm npe$ NSs 
are two nodeless r-modes 
(which behave as normal and superfluid modes at a given temperature), 
while $\rm npe\mu$ NSs harbours, 
in addition, 
an infinite set of $r_n$-modes with nodes 
(here $n \geq 1$ is the number of nodes of $r$-mode, see section \ref{resII}). 
These modes are of purely superfluid nature 
(counter-motion dominates, 
see section \ref{resII} and Fig.\ \ref{Fig:efnpemu}), 
and hence we will denote them $r^s_n$-modes.
It is 
quite natural to assume
(although we have not checked this by direct calculation)
that these $r^s_n$-modes 
correspond to the superfluid inertial $i^s_0$-modes 
of $\rm npe$ NSs (in analogy with non-superfluid NSs discussed above). 

So, why do the superfluid (mixed) inertial $i^s_0$-modes of $\rm npe$ NSs 
mutate into
the superfluid $r^s_n$-modes under stratification by muons?
Let us remind that for the modes with $\sigma \sim \Omega$ 
the electron and muon continuity equations 
imply that $\xi_r \rightarrow 0$ at $\Omega \rightarrow 0$ 
(see section \ref{diff}), 
which means that $\xi_{{\rm b}r}=z_r$ 
at $\Omega \rightarrow 0$.
Note also that
there is a clear 
distinction 
between the normal and superfluid modes 
due to a very good decoupling 
of superfluid and normal degrees of freedom 
(see \citealt{hap09,gk11,kg12,dg16}).
In the limiting case of perfect decoupling 
the superfluid modes cannot excite baryon displacements $\xi_{\rm b}$, 
and thus can meet the condition $\xi_r=0$ 
only by vanishing $z_r$ 
(i.e., transforming into purely toroidal modes). 
Our numerical results indicate 
that in the situation of approximate decoupling 
this conclusion remains the same.

Moreover, our results imply that stratification by muons 
does not transform normal (mixed) inertial modes $i_0$ 
into normal r-modes with the nodes 
(we do not find such modes in the calculated spectrum).
We expect that they survive in $\rm npe\mu$ NSs.
This assumption seems to be reasonable, 
since stratification by muons put only one constraint, 
$\xi_r=0$ at $\Omega \rightarrow 0$, limiting the variety of (mixed) inertial modes, 
but not 
excluding 
them completely.
Why do the normal inertial modes 
can retain their nature 
(in contrast to superfluid modes) 
under stratification by muons? 
Let us again illustrate our point by assuming perfect decoupling 
of superfluid and normal degrees of freedom.
Perfect decoupling
does not mean that normal modes 
cannot excite the superfluid displacement
${\pmb z}$ (see \citealt{gk11,kg12,gkcg13,dg16}). 
Thus, the condition $\xi_r=0$ can be 
easily satisfied for normal inertial modes 
by choosing $z_r$ such that $\xi_r= \xi_{{\rm b}r}-z_r=0$
at $\Omega \rightarrow 0$.

\begin{figure}
    \begin{center}
        \leavevmode
        \epsfxsize= 6.0in \epsfbox{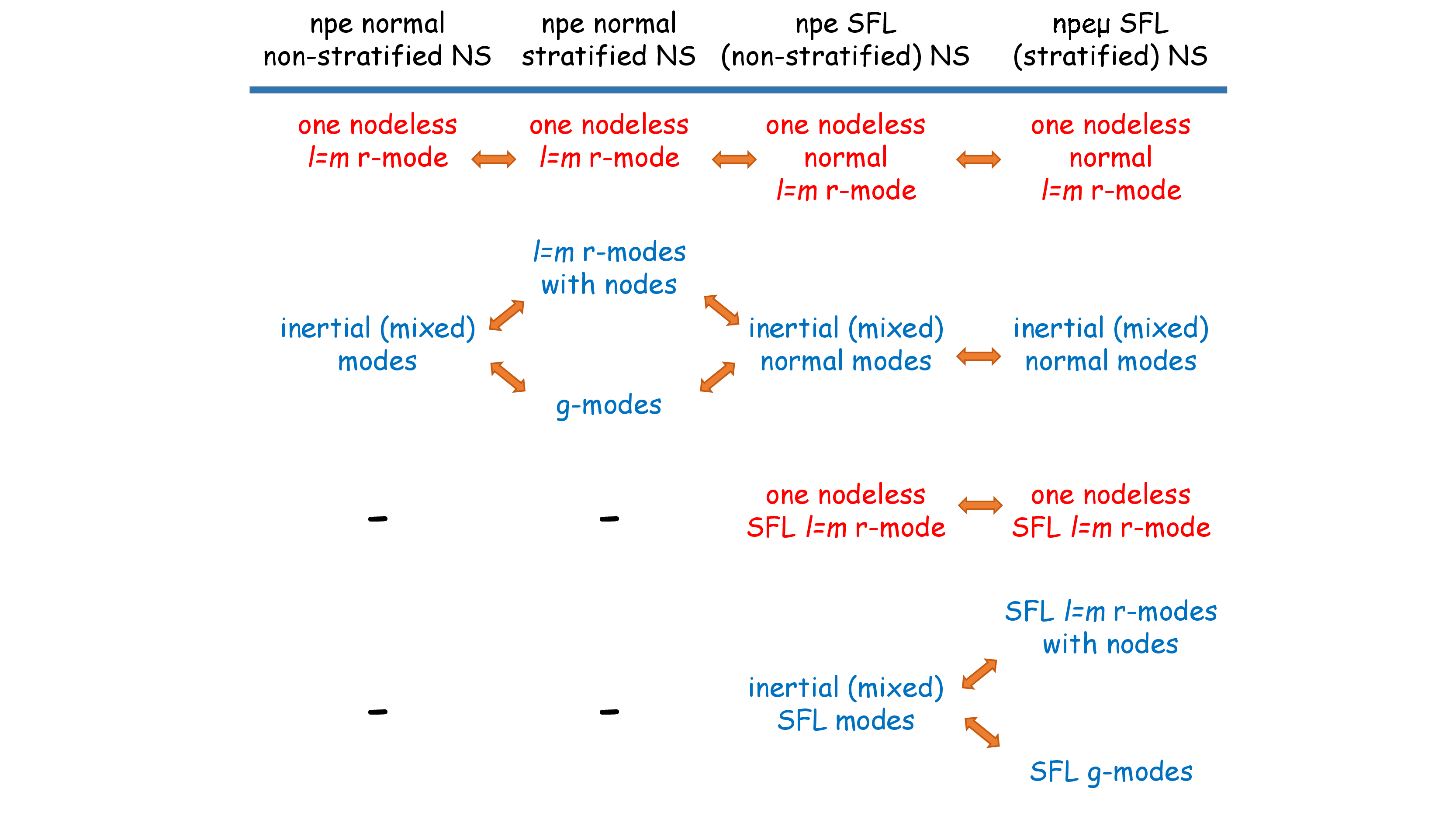}
    \end{center}
    \caption{The correspondence scheme for the modes with $\sigma \propto \Omega$. The case $Y_{\rm np}=0$ is considered. 
		Mixed (poloidal and toroidal) inertial modes, toroidal r-modes with $l=m$ and g-modes in non-stratified and stratified NSs are shown. 
		The r-modes with $l\neq m$, which are present in non-superfluid NSs are not included (they have $\sigma \rightarrow 0$ at $\Omega \rightarrow \Omega_0$ and are not interesting for us).}
    \label{Fig:modes}
\end{figure}

By analogy with the non-superfluid star, 
it seems reasonable
that, at low rotation rates
(when the spin frequency 
is smaller than the corresponding g-mode frequency of a non-rotating NS, 
that is hundreds of Hz for superfluid $\rm npe\mu$ NSs, see \citealt{kg14,pah16}), 
stratification by muons transforms $i^s_k$-modes of $\rm npe$ NSs (with $k\geq 1$, see footnote \ref{foot_k} for the definition of $k$)
into g-modes 
(they have finite $\sigma$ at $\Omega \rightarrow 0$ 
and hence are not required to have 
$\xi_r=0$ at $\Omega \rightarrow 0$). 
At the same time, normal $i_k$ ($k\geq 1$) modes 
do not transform into g-modes, 
retaining their nature. 
The scheme, illustrating this complex behavior, is presented in Fig. \ref{Fig:modes}.

In accordance with the numerical results of \cite{ly03}, 
where SFL $\rm npe$ NSs were considered, 
it is 
$i^s_0$-modes (see footnote \ref{foot_k})
that
experience avoided-crossings 
with normal r-mode in $\rm npe$ NSs. 
As has already been discussed above, 
stratification by muons transforms 
$i^s_0$-modes into $r_n^s$-modes, 
studied here.
In contrast,
superfluid $i^s_k$-modes ($k\geq 1$) 
have rather high eigenfrequencies and do not meet 
the normal r-mode (\citealt{ly03}) in $\rm npe$ NSs. 
Stratification by muons 
mixes $i^s_k$ modes ($k\geq 1$)
with g-modes and thus
(additionally) 
increases
their
eigenfrequencies 
to the corresponding g-mode eigenfrequencies 
in the limit $\Omega\rightarrow 0$.
Consequently, $i^s_k$ modes ($k\geq 1$) 
should not interact (form avoided crossings) 
with normal r-mode in $\rm npe\mu$ NSs as well.

To sum up, the $r^s_n$-modes ($n \geq 1$) of $\rm npe\mu$ NSs calculated in this paper, 
being the counterparts of $i^s_0$-modes 
of $\rm npe$ NSs
calculated in \cite{ly03},
were exactly the modes 
interesting for us since we were looking for
the resonance interaction 
(avoided crossings)
of normal r-mode with superfluid modes. 
Note that \cite{ly03} does not discuss explicitly
the interaction of superfluid nodeless r-mode 
with the normal r-mode.
However, their results imply that 
it takes place 
in NSs with weak entrainment
(when $Y_{\rm np}$ is small;
see also section \ref{res} and the work by \citealt{gck14b}).

In this paper we assumed that $Y_{\rm np}=0$. 
This limit implies $h(x)=h_1(x)$ 
(see equations \ref{beta}--\ref{y}) 
and it makes possible purely toroidal superfluid modes.
In this simplified model we 
found
that all purely toroidal 
modes have the same eigenfrequency at the leading order in rotation, 
$\sigma_0=\frac{2m}{l(l+1)}$, 
and only modes with $l=m$ exist. 
Next order 
approximation
removes this degeneracy 
(see Figs. \ref{Fig:spectrumnpe} and \ref{Fig:spectrumnpemu}).
As a result, 
normal r-mode exhibits
avoided-crossings with the nodeless superfluid mode, 
while other $r^s_n$ modes (with $n\geq 1$) have substantially lower $\sigma_1$
and do not interact with the normal r-mode.

If one relaxes the assumption $Y_{\rm np}=0$
\footnote{More rigorously, if one relaxes the assumption that $h_1(x)/h(x)$ is independent of $x$.},
one does not find {\it global} eigenfrequencies for
purely toroidal superfluid modes
in the leading order 
in $\Omega/\Omega_0$ 
(see the corresponding discussion in section \ref{equations} and in \citealt{agh09}). 
What one finds, instead, is 
$x$-dependent eigenfrequency 
which should be regularized to some global value 
by admixing poloidal component to the solution, 
that is,
by making superfluid radial displacement 
$z_r$ non-zero at $\Omega \rightarrow 0$.
Still, 
the NS eigenmodes 
should vary continuously with 
variation of $Y_{\rm np}$.
Thus, the limit $Y_{\rm np}=0$ should be a good approximation 
for small but non-zero $Y_{\rm np}$. 
How small should it be? The answer is not so straightforward.

At first glance, 
the real eigenfrequency should be of the order of 
the typical value of $\frac{2m}{l(l+1)}\frac{h_1(x)}{h(x)}$. 
While the variation of $\frac{2m}{l(l+1)}\frac{h_1(x)}{h(x)}$ 
(let us denote it $\delta \sigma_0$) 
throughout the star is small, 
the required admixture of the poloidal component 
should also be
small, 
not affecting oscillations substantially. 
It seems that if $\sigma_1\Omega^2>\delta \sigma_0/\sigma_0$, 
the effect on eigenmodes due 
to second order rotational corrections 
should be more pronounced than the effect of non-zero $Y_{\rm np}$.

However, the situation may be not so simple. 
Let us remind that, 
due to the condition 
$\xi_r \rightarrow 0$ at $\Omega \rightarrow 0$ 
(valid for $\rm npe\mu$-matter, see section \ref{diff}),
$\xi_{{\rm b}r}=z_r$ 
at $\Omega \rightarrow 0$.
Thus, the non-zero value of $Y_{\rm np}$ 
leads to the excitation of $\xi_{{\rm b}r}$ in superfluid modes. 
On the other hand, superfluid and normal modes 
are decoupled 
to a
good accuracy (\citealt{gk11,kg12,dg16}). 
In the limiting case of perfect decoupling 
this means that superfluid modes cannot excite 
baryon displacements ${\pmb \xi}_{\rm b}$. 
Thus, in this limiting case superfluid modes 
with $\sigma \sim \Omega$ 
should
be 
completely pushed out of that part of $\rm npe\mu$ region 
where
$Y_{\rm np}\neq 0$ 
(where protons are superconducting).
In reality, the decoupling is not perfect 
and ${\pmb \xi}_{\rm b}$ can be excited, to some extent, 
by superfluid modes 
in $\rm npe\mu$ region.
However, 
the above consideration clearly indicates 
that decoupling leads to at least partial expulsion 
of the eigenfunctions from the $\rm npe\mu$ region 
at non-zero $Y_{\rm np}$. 
Consequently, even small value of $Y_{\rm np}$ 
can, in principle, substantially modify 
the superfluid eigenmodes if decoupling is good. 
All this complicated behaviour is driven by the interplay 
of three small parameters: rotation frequency, $Y_{\rm np}$, 
and decoupling parameter (see \citealt{gk11} 
for the definition and more details on this parameter).
 
The above discussion clearly indicates 
that stratification by muons 
plays a key role in
defining the low-frequency spectrum of superfluid NSs,
whose rotation frequencies are smaller or comparable to the g-mode frequencies.
Keeping in mind that typical g-mode frequencies 
in superfluid $\rm npe\mu$ NSs 
can reach 
hundreds of Hz (see \citealt{kg14,dg16,pah16}),
account for muons appears to be crucial 
even for the most rapidly rotating NSs.

Summarizing, 
using the approximation of vanishing entrainment
between the 	
superfluid neutrons and superconducting protons, 
we calculated the spectrum of toroidal modes of slowly rotating superfluid neutron stars.
We analysed NSs with both $\rm npe$ and $\rm npe\mu$ compositions of their core.
We found that $\rm npe$ NSs harbour only two types of toroidal modes 
--- nodeless normal and nodeless superfluid r-modes.
In contrast, 
$\rm npe\mu$ NSs additionally possess an infinite set of superfluid r-modes with nodes. 
For both NS compositions, 
normal r-mode exhibits avoided-crossings 
with the nodeless superfluid r-mode.
Other superfluid r-modes in $\rm npe\mu$ NSs have lower frequencies 
and do not 
approach (and form avoided crossings with)
the normal r-mode. 
In addition, we  calculated the `instability windows' for r-modes, 
accounting for the mode dissipation due to mutual friction and shear viscosity. 
We found that the instability window is split up by two stability peaks 
so that all the observed sources can be interpreted 
as either lying inside the stability region 
or climbing the left edge of one of those peaks.
This result confirms 
the phenomenological scenario of \cite{gck14a,gck14b,cgk14,cgk17} 
which was suggested to explain stability of most rapidly rotating NSs.
In the subsequent work we plan to relax the simplifying assumption $Y_{\rm np}=0$,
adopted in the present paper, 
and to accurately calculate the temperature-dependent spectrum of inertial modes
in rotating superfluid NSs.

\section*{Acknowledgments}
We are grateful to Andrey Chugunov and Vasiliy Dommes for valuable comments and discussions.

\appendix
\section{Derivation of equations of section 2.2}
\label{derivation}

We start with $r$ component of Euler equation (\ref{euler})
\begin{eqnarray}
2\imath \sigma \Omega {\rm sin}\theta \xi_{\rm b \phi}=-\frac{\delta w}{w_0^2}\frac{\partial P_0}{\partial r}+\frac{1}{w_0}\frac{\partial \delta P}{\partial r}. \label{eulerrapp}
\end{eqnarray}
Making use of 
\begin{eqnarray}
{\pmb \nabla} w_0=\frac{1}{c^2}\left({\pmb \nabla} P_0+\mu_{\rm n}{\pmb \nabla} n_{\rm b0}\right)=\frac{1}{c^2}\left(1+\mu_{\rm n}\frac{\partial n_{\rm b}}{\partial P}\right){\pmb \nabla} P_0, \label{thermod}
\end{eqnarray}
that follows from 
the formula $\delta \epsilon =\mu_{\rm n} \delta n_{\rm b}$, 
and the fact that in the equilibrium $\Delta \mu_{\rm l}=0$,
(\ref{eulerrapp}) can be rewritten as
\begin{eqnarray}
2\imath \sigma \Omega {\rm sin}\theta \xi_{\rm b \phi}=\frac{\partial }{\partial r}\frac{\delta P}{w_0}-\frac{\mu_{\rm n}}{w_0^2 c^2}\left(\frac{\partial n_{\rm b}}{\partial \Delta \mu_{\rm e}}\Delta \mu_{\rm e}+\frac{\partial n_{\rm b}}{\partial \Delta \mu_{\rm \mu}}\Delta \mu_{\rm \mu} \right) \frac{\partial P_0}{\partial r}. \label{Eulerr}
\end{eqnarray}

$\phi$ component of equation (\ref{euler}) gives us $\delta P$,
\begin{eqnarray}
\delta P=-\frac{\imath}{m} a_\phi w_0 r \,{\rm sin}\theta,  \label{deltaPapp1}
\end{eqnarray}
where we introduced the vector 
${\pmb a}\equiv \sigma^2 {\pmb \xi}_{\rm b}-2\imath \sigma {\pmb \Omega} \times {\pmb \xi}_{\rm b}$.

$\theta$-component of the Euler equation is 
\begin{eqnarray}
a_\theta=-\frac{\delta w}{w_0^2r}\frac{\partial P_0}{\partial \theta}+\frac{1}{w_0 r}\frac{\partial \delta P}{\partial \theta}. \label{thetaEuler1}
\end{eqnarray}
We will be interested 
in the first two terms 
in the 
expansion of the right-hand side of this equation 
in small parameter $\Omega/\Omega_0$.
Thus, the independent variables $r$, $\theta$, $\phi$ 
should be changed to $x$, $\theta$, $\phi$ more cautiously.
Applying equation (\ref{changevar}) to equation (\ref{thetaEuler1}), we obtain
\begin{eqnarray}
a_\theta=\frac{1}{w_0 r}\frac{\partial \delta P}{\partial \theta}-2\Omega^2 \alpha(x) {\rm cos} \theta {\rm sin}\theta\left(-\frac{\delta w}{w_0^2}\frac{\partial P_0}{\partial x}+\frac{1}{w_0}\frac{\partial \delta P}{\partial x}\right).
\label{a5}
\end{eqnarray}
Making use of equation (\ref{eulerrapp}) and substituting the expression (\ref{deltaPapp1}) 
into (\ref{a5}), we get 
(note that here $x$, $\theta$, and $\phi$ are independent variables)
\begin{eqnarray}
a_\theta=-\frac{\imath}{w_0 r m}\frac{\partial}{\partial \theta}\left( a_\phi w_0 r \,{\rm sin}\theta\right)-2\Omega^2 \alpha(x) {\rm cos} \theta {\rm sin}^2\theta\left(2\imath\sigma\Omega\xi_{{\rm b}\phi}\right).
\end{eqnarray}
Substituting now the expression (\ref{shape}) for $r$, 
we arrive at
\begin{eqnarray}
a_\theta= -\frac{\imath}{m}\frac{\partial}{\partial \theta}{\rm sin}\theta a_\phi-\frac{\imath}{m}2\Omega^2 \alpha(x) {\rm cos}\theta{\rm sin}^2 \theta \left(a_\phi+2\sigma\Omega m \xi_{{\rm b}\phi} \right).
\end{eqnarray}
This equation, written to the leading order in $\Omega/\Omega_0$, 
gives us equation (\ref{Euler0});
the next-to-leading order in $\Omega/\Omega_0$ gives equation (VII).
Proceeding in a similar way with equation (\ref{sfl}), 
one finds equations (\ref{sfl0general}) (with $h=h_1$) and (VIII).

\section{npe-matter: Analytic solution}
\label{npe}

Here we present the oscillation
equations and discuss their solution
for NSs, 
whose cores are composed of npe-matter.
To the leading order in $\Omega/\Omega_0$
the oscillation equations 
(equations \ref{contb0}, \ref{contl0}, \ref{Euler0}, and \ref{sfl0general}) 
are the same as for $\rm npe\mu$-matter.
However, 
in the next order
the system of equations
for npe-matter will differ from $\rm npe\mu$-equations (I)--(VIII)
and consist of:
\\
(I) the continuity equation for baryons:
\begin{eqnarray}
\frac{1}{n_{\rm b}} \left(\frac{\partial n_{\rm b}}{\partial P} \delta P_1+\frac{\partial n_{\rm b}}{\partial \Delta \mu_{\rm e}} \Delta \mu_{{\rm e}1}\right)=\nonumber \\ 
-\frac{1}{x^2 n_{\rm b}}\frac{\partial}{\partial x}(x^2 n_{\rm b} \xi_{{\rm b}r}^1)-\frac{1}{x {\rm sin} \theta}\left(\frac{\partial}{\partial \theta}{\rm sin}\theta \xi_{\rm b \theta}^1+\imath m\xi_{\rm b\phi}^1\right)-2\alpha(x){\rm cos}\theta{\rm sin}\theta\left(g\frac{w_0}{n_{\rm b}}\frac{\partial n_{\rm b}}{\partial P}-\frac{\partial}{\partial x}\right)\xi_{\rm b \theta}^0, \label{contb2app}
\end{eqnarray}
where the independent variables in thermodynamic derivatives are $P$ 
and $\Delta \mu_{\rm e}=\mu_{\rm n}-\mu_{\rm p}-\mu_{\rm e}$.
\\
(II) the continuity equation for electrons:
\begin{eqnarray}
\frac{1}{n_{\rm e}}\left(\frac{\partial n_{\rm e}}{\partial P} \delta P_1+\frac{\partial n_{\rm e}}{\partial \Delta \mu_{\rm e}} \Delta \mu_{{\rm e}1}\right)=\nonumber \\ 
-\frac{1}{n_{\rm e}x^2}\frac{\partial}{\partial x}(x^2 n_{\rm e} \xi_{{\rm b}r}^1)-\frac{1}{x {\rm sin} \theta}\left(\frac{\partial}{\partial \theta}{\rm sin}\theta \xi_{{\rm b}\theta}^1+\imath m\xi_{{\rm b}\phi}^1\right)-2\alpha(x){\rm cos}\theta{\rm sin}\theta \left(g\frac{w_0}{n_{\rm e}}\frac{\partial n_{\rm e}}{\partial P}-\frac{\partial}{\partial x}\right)\xi_{{\rm b}\theta}^0 \nonumber \\
+\frac{1}{n_{\rm e}x^2}\frac{\partial}{\partial x}\left(x^2 n_{\rm e} z_r^1\right)+\frac{1}{x {\rm sin} \theta}\left(\frac{\partial}{\partial \theta}{\rm sin}\theta z_{\theta}^1+\imath m z_{\phi}^1\right)+2\alpha(x){\rm cos}\theta{\rm sin}\theta \left(g\frac{w_0}{n_{\rm e}}\frac{\partial n_{\rm e}}{\partial P}-\frac{\partial}{\partial x}\right)z_{\theta}^0, \label{contl2app}
\end{eqnarray}
(III) $r$-component of the Euler equation
\begin{eqnarray}
2\imath \sigma_0 {\rm sin}\theta \xi_{\rm b \phi}^0=\frac{\partial }{\partial x}\frac{\delta P_1}{w_0}-\frac{\mu_{\rm n}}{w_0^2 c^2}\left(\frac{\partial n_{\rm b}}{\partial \Delta \mu_{\rm e}}\Delta \mu_{{\rm e} 1}\right) \frac{\partial P_0}{\partial x}, \label{Eulerrapp}
\end{eqnarray}
(IV) $\phi$-component of the Euler equation
\begin{eqnarray}
\sigma_0^2 \xi_{{\rm b}\phi}^0-2\imath \sigma_0 {\rm cos}\theta \xi_{{\rm b}\theta}^0=\frac{\imath m}{x {\rm sin}\theta w_0}\delta P_1, \label{deltaPapp}
\end{eqnarray}
(V) $r$-component of the superfluid equation
\begin{eqnarray}
2\imath h  \sigma_0 {\rm sin}\theta z_{\phi}^0=c^2 n_{\rm e} \frac{\partial \Delta \mu_{{\rm e} 1}}{\partial r}, \label{sflrapp} 
\end{eqnarray}
(VI) $\phi$-component of the superfluid equation
\begin{eqnarray}
\sigma_0^2 z_{\phi}^0-2\imath \sigma_0 {\rm cos}\theta z_{\theta}^0=c^2\frac{\imath m}{x {\rm sin}\theta h}n_{\rm e} \Delta \mu_{{\rm e} 1}.  \label{deltamuapp} 
\end{eqnarray}
These equations should be supplemented with $\theta$-components 
of the Euler and superfluid equations, which coincide 
with the corresponding  equations in $\rm npe\mu$-matter
[see equations (VII) and (VIII) in the main text].

The analysis of this system of equations ensures that, as in the $\rm npe\mu$-case,
only 
modes with $l=m$ are possible in $\rm npe$-matter (we remind that this system is valid for toroidal modes only).
To solve this system, 
let us express $\Delta \mu_{{\rm e}1}$ (defined by equation \ref{deltaf}) 
from equation (\ref{deltamuapp}) 
and substitute 
the answer
into equation (\ref{sflrapp}).
The solution to the resulting equation is remarkably simple, 
\begin{eqnarray}
C_{z}(x)=\frac{n_{\rm e}(x)\,x^m}{h(x)},\label{Czlm}
\end{eqnarray}
where we set to unity the dimensional integration constant.
Expressing now $\delta P_1$ from equation (\ref{deltaPapp}) 
and substituting $\delta P_1$ and $\Delta \mu_{{\rm e}1}$ into equation (\ref{Eulerrapp}), 
we obtain the first-order inhomogeneous differential equation for $C(x)$,
\begin{eqnarray}
C'(x)=\frac{m C(x)}{x}+\frac{x^m \partial n_{\rm b}/\partial \Delta\mu_{\rm e}(x) \mu_{\rm n}(x) P_0'(x)}{c^4 w_0^2(x)}, \label{Cshtrich}
\end{eqnarray}  
whose solution is
\begin{eqnarray}
C(x)=x^m\left(C_{0}+\int_0^x\frac{\partial n_{\rm b}/\partial \Delta \mu_{\rm e}(x_1)\mu_{\rm n}(x_1)P_0'(x_1)}{c^4 w_0^2(x_1)}dx_1\right),\label{Clm}
\end{eqnarray}
where $C_{0}$ is an integration constant to be determined below.

Substituting now the expansions 
(\ref{expxib})--(\ref{expRz}) into (\ref{Eulertheta}) 
divided by ${\rm sin}\theta$, 
and equating coefficients 
at the terms proportional to $P_{m}^m$, 
we obtain an algebraic equation relating $Q^1(x)$ to $\xi_{{\rm b} r}^1(x)$ 
and $C(x)$ 
(note that the function $T^1(x)$ cancels out
in the terms proportional to $P_{m}^m$). 
Similarly, using
equation (\ref{sfltheta}), 
we relate $Q_{z}^1(x)$ to $z_{r}^1(x)$ and $C_{z}(x)$.

Now we substitute the derived expression for $Q^1(x)$, 
as well as (\ref{Cshtrich}), 
into the continuity equation for baryons (\ref{contb2app}). 
Equating then the coefficients at $P_{m+1}^m$ 
and using the expressions (\ref{Clm}) and (\ref{Czlm}), 
we obtain the following first-order inhomogeneous differential equation 
for $\xi_{{\rm b} r}^1(x)$, 
\begin{eqnarray}
\frac{d}{dx}\xi_{{\rm b} r}^1(x)+A(x)\xi_{{\rm b} r}^1(x)-B_{00}(x)-C_{0}B_{01}(x)-\sigma_1 B_{10}(x)-\sigma_1 C_{0} B_{11}(x)=0,
\end{eqnarray}
where $A(x)$, $B_{00}(x)$, $B_{10}(x)$, $B_{01}(x)$, and $B_{11}(x)$ are certain functions of $x$ defined by equation (\ref{contb2app}).
The solution to this equation is 
\begin{eqnarray}
\xi_{{\rm b} r}^1(x)=H(x)\left(\xi_0+\int_0^x \frac{B_{00}(x_1)}{H(x_1)}dx_1+C_{0} \int_0^x \frac{B_{01}(x_1)}{H(x_1)}dx_1+\sigma_1 \int_0^x \frac{B_{10}(x_1)}{H(x_1)}dx_1+\sigma_1 C_{0}\int_0^x \frac{B_{11}(x_1)}{H(x_1)}dx_1 \right),
\end{eqnarray}
where $\xi_0$ is the integration constant and
\begin{eqnarray}
H(x)={\rm exp}\left(-\int A(x)dx\right)=\frac{1}{n_{\rm b}(x)x^{m+3}}.  
\end{eqnarray}
Following the same strategy with 
the electron continuity equation (\ref{contl2app}), 
we find for $z_{r}^1(x)$,
\begin{eqnarray}
z_{r}^1(x)=H_z(x)\left(z_0+\int_{x_{\rm sfl1}}^x \frac{B_{z00}(x_1)}{H_z(x_1)}dx_1+C_{0} \int_{x_{\rm sfl1}}^x \frac{B_{z01}(x_1)}{H_z(x_1)}dx_1+\sigma_1 \int_{x_{\rm sfl1}}^x \frac{B_{z10}(x_1)}{H_z(x_1)}dx_1\right),
\end{eqnarray}
where $x_{\rm sfl1}$ corresponds to 
the superfluid/nonsuperfluid 
inner boundary, 
$z_0$ is the integration constant and
\begin{eqnarray}
H_z(x)={\rm exp}\left(-\int A_z(x)dx\right)=\frac{1}{n_{\rm e}(x)x^{m+3}}.  
\end{eqnarray}
Analysis of the hydrodynamic equations in the stellar centre implies that 
$\xi_{{\rm b} r}^1(x)\propto x^m$, 
$z_{r}^1(x)\propto x^m$ at $x\rightarrow 0$, 
and thus $\xi_0=0$ and $z_0=0$. 
If the stellar centre is non-superfluid then $z_0=0$ anyway,
because
at the superfluid/non-superfluid interfaces 
$z_{r}^1=0$ (see section \ref{bc}),
and application of this condition to 
the inner 
boundary of the superfluid region 
($x=x_{\rm sfl1}$) gives us $z_0=0$.
The finiteness of the function $\xi_{{\rm b} r}^1(x)$ at the surface ($x=R$)
and vanishing of $z_{r}^1(x)$ at the 
outer
superfluid/non-superfluid 
boundary
($x={x_{\rm sfl2}}$) imply 
\begin{eqnarray}
\int_0^R \frac{B_{00}(x_1)}{H(x_1)}dx_1+C_{0} \int_0^R \frac{B_{01}(x_1)}{H(x_1)}dx_1+\sigma_1 \int_0^R \frac{B_{10}(x_1)}{H(x_1)}dx_1+\sigma_1 C_{0}\int_0^R \frac{B_{11}(x_1)}{H(x_1)}dx_1=0, \label{lin1}\\
\int_{x_{\rm sfl1}}^{x_{\rm sfl2}} \frac{B_{z00}(x_1)}{H_z(x_1)}dx_1+C_{0} \int_{x_{\rm sfl1}}^{x_{\rm sfl2}} \frac{B_{z01}(x_1)}{H_z(x_1)}dx_1+\sigma_1 \int_{x_{\rm sfl1}}^{x_{\rm sfl2}} \frac{B_{z10}(x_1)}{H_z(x_1)}dx_1=0,   \label{lin2}
\end{eqnarray}
The system of equations (\ref{lin1})--(\ref{lin2}) 
for $\sigma_1$ and $C_{0}$ has two solutions corresponding to two oscillation modes
discussed in the text.

\section{Relating mutual friction coefficient $\beta_{\rm MF}$ to the relaxation time $\tau_{\rm v}$ at finite temperature}
\label{mf}

To relate the mutual friction coefficient $\beta_{\rm MF}$ (or $B$) to the microphysics parameters, 
it is customary to use 
(e.g., \citealt{mendell91,asc06}) the notion of the relaxation time $\tau_{\rm v}$ (\citealt{feibelman71,sss82,als84}).
This time characterizes relaxation 
of the electron velocity, $\pmb{u}$
(which is also the velocity of other particle species, 
coupled to electrons on time-scales much shorter than $\tau_{\rm v}$, 
i.e., muons, protons, and neutron thermal excitations; 
hereafter normal liquid component),
to the velocity of the static (fixed) neutron vortices. 
Note that only the component of $\pmb{u}$ orthogonal to the direction of vortices
experiences relaxation;
let us denote it $\pmb{u}_{\perp}$. 
In the reference frame of fixed vortices (where their velocity ${\pmb V}_{\rm L}=0$) the dissipative force acting on the flow of normal component is 
\begin{eqnarray}
{\pmb F}=-
\frac{\mu_{\rm n} n_{\rm b}-\mu_{\rm n} \mu_k Y_{{\rm n}k}}{c^2}\frac{{\pmb u_\perp}}{\tau_{\rm v}},
\end{eqnarray}
Here $(\mu_{\rm n} n_{\rm b}-\mu_{\rm n} \mu_k Y_{{\rm n}k})/c^2$ represents the density of normal component  (\citealt{ga06}).
In an arbitrary frame, if we relax the assumption about static vortices, 
the friction force on the normal component is
\begin{eqnarray}
{\pmb F}=-\frac{\mu_{\rm n} n_{\rm b}-\mu_{\rm n} \mu_k Y_{{\rm n}k}}{c^2}\,\frac{{\pmb u}-{\pmb V}_{\rm L}}{\tau_{\rm v}}.
\end{eqnarray}
The same dissipative force, taken with the opposite sign, acts on neutron vortices.
Since the total force on vortices should vanish (they are practically massless), 
this dissipative force must be 
compensated by the Magnus force acting on vortices from the neutron superfluid.
Consequently, we can conclude that vortices 
act on the neutron superfluid with the (minus) Magnus force, 
or with
\begin{eqnarray}
-{\pmb F}=\frac{\mu_{\rm n} n_{\rm b}-\mu_{\rm n} \mu_k Y_{{\rm n}k}}{c^2}\,\frac{{\pmb u}-{\pmb V}_{\rm L}}{\tau_{\rm v}}
\end{eqnarray}
Consider now the weak-drag limit relevant for NS matter (\citealt{mendell91,asc06}). 
In this limit the vortex velocity is 
(see equation 101 in \citealt{gd16}, 
where we skipped the term $\propto \beta$, since it is small)
\begin{eqnarray}
 V_{({\rm L})}^\mu=u^\mu-\mu_{\rm n} \alpha Y_{{\rm n}k}w_{(k)\,\nu} \perp^{\mu \nu},
\end{eqnarray}
and $\alpha=-1/(\mu_{\rm n} \mu_k Y_{{\rm n}k})$. 
In the locally flat comoving 
reference frame [where $u^\mu=(1,0,0,0)$] 
the spatial components of $V_{\rm (L)}^\mu$ are 
\begin{eqnarray}
{\pmb V}_{\rm L}=-\mu_{\rm n} \alpha Y_{{\rm n}k}{\pmb w}_{(k)},
\end{eqnarray}
The dissipative force on the neutron superfluid is then
\begin{eqnarray}
-\pmb F=\frac{\mu_{\rm n} n_{\rm b}-\mu_{\rm n} \mu_k Y_{{\rm n}k}}{c^2}\,\frac{\mu_{\rm n} \alpha Y_{{\rm n}k}\pmb w_{(k)}}{\tau_{\rm v}}. \label{f_diss}
\end{eqnarray}
On the other hand, this force can be deduced from the equation (\ref{f_gen}). 
Using equation (I7) of \cite{gd16}, 
one can show that the force on the neutron superfluid equals 
$-\pmb F=-n_{\rm n} \pmb f
\mu_{\rm n} \mu_k Y_{{\rm n}k}/c^2$, 
where ${\pmb f}$ is the spatial part of the four-vector $f^{\mu}$ 
(see equation 100 of \citealt{gd16}), 
taken
in the comoving frame. 
Using equation (100) from \cite{gd16},
the latter expression can be represented, 
in the 
comoving 
frame, as
\begin{eqnarray}
-\pmb F=-\frac{\mu_{\rm n} \mu_i Y_{{\rm n}i}}{c^2} \beta_{\rm MF}\mathcal{V}_{({\rm M})} \,Y_{{\rm n}k}\pmb w_{(k)}, \label{f_mf}
\end{eqnarray} 
where $\mathcal{V}_{({\rm M})}=2\Omega \mu_{\rm n}/c^2$ 
in the case of solid-body rotation. 
Comparing now (\ref{f_diss}) and (\ref{f_mf}), we obtain (in dimensional form)
\begin{eqnarray}
\beta_{\rm MF}=\frac{c^2}{2\Omega\tau_{\rm v}}\frac{(\mu_{\rm n} n_{\rm b}-\mu_{\rm n} \mu_k Y_{{\rm n}k})}{(\mu_{\rm n} \mu_k Y_{{\rm n}k})^2}. 
\end{eqnarray}
%

\section{Typos in the paper by Provost et al. (1981)}
\label{typos}

When considering the r-mode problem 
in superfluid NSs 
we followed the approach of \cite{pbr81}. 
Although generally this paper is extremely 
illuminating,
it
contains a number of typos 
and some incorrect statements in its Appendix.
Here we would like to list them. 
First of all, the definition of $g$ 
[between the formulas (7d) and (8)] 
reads $g=\rho^{-1}\left|\frac{dp}{dx}\right|$. 
Next, the first equation in 
(10) should read
\begin{equation}
\xi_\phi^1=\frac{\imath m Q^1(x,\theta)}{{\rm sin}\theta}-\frac{d}{d \theta}R(x,\theta).
\end{equation}
Let us now turn to the Appendix of \cite{pbr81}, 
discussing barotropic stars.
We note that:

(i) the corrections $\xi^1_\theta$ 
and $\xi^1_\phi$ 
do not have to be purely poloidal;
however, the toroidal component 
does not appear in 
the lowest harmonics in equations (7c) and (7d). 

(ii)
In the second term in equation (A4)
$m+3$ should be replaced with $(m+3)/x$. 

(iii) In the solution (A5) 
$1/\gamma$ in the exponents should be replaced with $-1/\gamma$. 

(iv) Finally, we believe that 
the discussion 
of the behaviour of the solution near the surface is not correct.
The integral entering (A5) with $x^{-m-3}p^{-1/\gamma}(x)$ 
factored out (let us denote it by $\int^x_0 \ldots$) 
is regular near the stellar surface (since $1/\gamma>1$),
while the factor $1/p^{1/\gamma}(x)$ diverges. The only way to have a regular solution at the surface ($x=R$)
is to require $\int_0^R \ldots=0$. 
This condition 
determines
the eigenvalues $\sigma_1$.%
\footnote{Similar approach 
	is used in appendix \ref{npe} 
	to calculate the spectrum of superfluid $\rm npe$ NSs.}

We checked this interpretation 
and found that it is correct.
Namely,
we calculated the values of $\sigma_1$ in the Cowling approximation 
in the case of a polytropic star 
(polytropic index $n=1$). 
We verified that our results are 
in agreement with 
those
listed in the last column 
of Table I in \cite{ak01}. 
This table presents numerical values for the parameter $\omega_2$,
defined as
\begin{equation}
\sigma=\sigma_0\Omega+\omega_2\frac{\Omega^3 R^3}{G M} 
\label{omega2}
\end{equation}
and related to $\sigma_1$ 
by the expression: $\sigma_1=\omega_2/\sigma_0$. 
Note that, actually, \cite{ak01} give different definition for $\omega_2$ 
(see their equation 32). 
However, comparing the first and third columns 
of Table I in \cite{ak01} 
with Table I in \cite{lmo99} 
[in which the parameter $\kappa_2$ is 
presented, defined by equations (3.7) and (3.13)], 
one can check that, 
to generate Table I, 
\cite{ak01}
employed the definition (\ref{omega2}) for $\omega_2$. 
%


\label{lastpage}

\end{document}